\documentclass[aps,amsmath,onecolumn, showpacs,superscriptaddress,pre,10.5pt,nofootinbib]{revtex4-1}
\usepackage{amssymb}
\usepackage{amsmath}
\usepackage{graphicx}
\usepackage{array}
\usepackage{physics}
\usepackage{mathtools}
\usepackage{amssymb}
\usepackage{verbatim}
\usepackage{color}
\usepackage{bm}
\usepackage{bbm}
\usepackage{amsthm}
\usepackage[normalem]{ulem}
\usepackage[titletoc,title]{appendix}

\usepackage{color}
\usepackage[dvipsnames]{xcolor}
\definecolor{Blue}{rgb}{0.00, 0.00, 1.00}
\definecolor{Red}{rgb}{1.00, 0.00, 0.00}
\definecolor{Green}{rgb}{0.00, 0.50, 0.00}

\usepackage[colorlinks=true, urlcolor=blue, anchorcolor=blue, citecolor=blue,filecolor=blue,linkcolor=blue,menucolor=blue]{hyperref}

\newcommand{\bea}{\begin{eqnarray}}
\newcommand{\eea}{\end{eqnarray}}
\newcommand{\be}{\begin{equation}}
\newcommand{\ee}{\end{equation}}
\newcommand{\nn}{\nonumber}
\newcommand{\bee}{\begin{equation*}}
\newcommand{\eee}{\end{equation*}}

\colorlet{Mycolor1}{green!10!orange!90!}

\def\XXint#1#2#3{{\setbox0=\hbox{$#1{#2#3}{\int}$}
     \vcenter{\hbox{$#2#3$}}\kern-.5\wd0}}

\begin{document}

\title{Condensation transition in large deviations of self-similar Gaussian processes with stochastic resetting}

\author{Naftali R. Smith}
\email{naftalismith@gmail.com}
\affiliation{Department of Solar Energy and Environmental Physics, Blaustein Institutes for Desert Research, Ben-Gurion University of the Negev, Sede Boqer Campus, 8499000, Israel}
\author{Satya N. Majumdar}
\email{satya.majumdar@universite-paris-saclay.fr}
\affiliation{Universit{\'e} Paris-Saclay, CNRS, LPTMS, 91405, Orsay, France}

\begin{abstract}

We study the fluctuations of the area $A(t)= \int_0^t x(\tau)\, d\tau$ under a self-similar Gaussian process (SGP) $x(\tau)$ with Hurst exponent $H>0$ (e.g., standard or fractional Brownian motion, or the random acceleration process) that stochastically resets to the origin at rate $r$. Typical fluctuations of $A(t)$ scale as $\sim \sqrt{t}$ for large $t$ and on this scale the distribution is Gaussian, as one would expect from the central limit theorem. Here our main focus is on atypically large fluctuations of $A(t)$. In the long-time limit $t\to\infty$, we find that the full distribution of the area takes the form $P_{r}\left(A|t\right)\sim\exp\left[-t^{\alpha}\Phi\left(A/t^{\beta}\right)\right]$ with anomalous exponents $\alpha=1/(2H+2)$ and $\beta = (2H+3)/(4H+4)$ in the regime of moderately large fluctuations, and a different anomalous scaling form 
$P_{r}\left(A|t\right)\sim\exp\left[-t\Psi\left(A/t^{\left(2H+3\right)/2}\right)\right]$ in the regime of very large fluctuations. The associated rate functions $\Phi(y)$ and $\Psi(w)$ depend on $H$ and are found exactly.  Remarkably, $\Phi(y)$ has a singularity that we interpret as a first-order dynamical condensation transition, while $\Psi(w)$ exhibits a second-order dynamical phase transition above which the number of resetting events ceases to be extensive. The parabolic behavior of $\Phi(y)$ around the origin $y=0$ correctly describes the typical, Gaussian fluctuations of $A(t)$. Despite these anomalous scalings, we find that all of the cumulants of the distribution $P_{r}\left(A|t\right)$ grow linearly in time, $\langle A^n\rangle_c\approx c_n \, t$, in the long-time limit. For the case of reset Brownian motion (corresponding to $H=1/2$), we develop a recursive scheme to calculate the coefficients $c_n$ exactly and use it to calculate the first 6 nonvanishing cumulants.

\end{abstract}


\maketitle

{%
	\hypersetup{linkcolor=black}
	\tableofcontents
}

\section{Introduction}

\subsection{Background}

One of the problems that is of fundamental importance in non-equilibrium statistical mechanics and probability theory is the study of fluctuations in stochastic systems.
One class of such systems that has attracted much interest, especially over the last decade, is stochastic processes that included resetting to some state (which is usually the initial state) \cite{MZ99,VAME10, EM1,EM2,MV13,EM14,KMSS14,GMS14, CM15,CS15,MSS15a,MSS15b,Meylahn15, MV16,MC16,EM16,PKE16,NG16, Reuveni16,RLSTG16,MMV17,PR17,HT17, MSM18, CS18,EM18,VM18,GGC18,MajumdarOshanin18, EM19, BKP19, KG19, MPCM19a,MPCM19b,MPCM19c,Gupta19,LD19,MM19,PDRK19, DH2019, MMS20, WCKMS21, SW21,SSIM21,  VCWMS22, SGS22, SG22},
see \cite{EMS20} for a recent review. 
Systems which stochastically reset have recently been realized in optical trap experiments and these
experiments have, in turn, led to new interesting theoretical questions \cite{FPSRR20, BBPMC20, FBPCM21}.
They exhibit several features of interest: They typically reach a nonequilibrium steady state, even if the reset-free process is not stationary. Additionally, the resetting can lead to a significant decrease in the first-passage times.
The simplest example is reset Brownian motion (RBM): Brownian motion $x(t)$ with diffusion coefficient $D$ and with resetting events that occur at random times. The resetting is a Poisson process with rate $r$, and at each resetting event the position of the particle is set back to the origin, $x=0$. The resetting confines the particle to the vicinity of the origin, so that the probability density function (PDF) of its position reaches a steady-state at long times given by \cite{EM1}
\be
p_{\text{st}}\left(x\right)=\frac{\alpha_{0}}{2}e^{-\alpha_{0}\left|x\right|}
\ee
where $\alpha_{0} = \sqrt{r/D}$ is the inverse of the typical length scale of the particle's diffusion between resetting events.

This paper builds on results from recent studies \cite{Meylahn15,HT17, DH2019} on the effect of the confinement (due to resetting) on the distribution $P(A|t)$ of additive (or dynamical) observables of the form
\be
\label{dynamicalObservable}
A(t) =\int_{0}^{t}u\left(x\left(\tau\right),\dot{x}\left(\tau\right)\right)d\tau\,,
\ee
where $x(\tau)$ is a stochastic process which stochastically resets at rate $r$, and $u(x,\dot{x})$ is an arbitrary function.
In a broad class of stochastic systems (with or without resetting), $A(t)$ converges, in the long-time limit $t\to \infty$, to its corresponding ensemble-average value as long as the system is ergodic (and therefore self-averaging). There will, however, be fluctuations from this behavior, which are interesting to quantify.
In many systems, fluctuations of $A(t)$ from its average value decay exponentially in time, as described by the ``usual'' large-deviation principle (LDP):
\be
\label{eq:DVscaling}
P\left(A|t\right)\sim e^{-tI\left(A/t\right)}, \quad t \to \infty,
\ee 
i.e., the limit $-\lim_{t\to\infty}\ln P\left(at^{\beta}|t\right)/t^{\alpha}=I\left(a\right)$ exists,
with the standard exponents $\alpha=\beta=1$ and with a
``rate function'' $I(a)$. 
There is a well-established theory (sometimes referred to as Donsker-Varadhan (DV) theory) for showing the existence of LDP's and for calculating and studying the rate function $I(a)$ \cite{Bray, Majumdar2007, MS2017, DonskerVaradhan,Ellis,T2009,Touchette2018}. 
Some generic properties of $I(a)$ can be found: it is nonnegative, convex, and vanishes when its argument $a = A/t$ equals its corresponding ensemble-average value.

DV theory was extended to stochastically resetting processes in Refs. \cite{Meylahn15,HT17, DH2019}.
Remarkably, it was found that the confinement (due to resetting) can sometimes induce an LDP of the standard type \eqref{eq:DVscaling} in the resetting process, even if the reset-free process does not satisfy this LDP.
Intriguingly, it was found there that for the apparently simple particular case of the area $A(t)$ under an RBM, the probability to observe a given value $A(t)/t$ decays slower than exponentially in $t$ at long times, i.e., the usual LDP \eqref{eq:DVscaling} holds trivially with a vanishing rate function $I(a)=0$. Therefore, Eq.~\eqref{eq:DVscaling} does not correctly capture the full distribution of $A(t)$, which has remained unknown. 
The scaling \eqref{eq:DVscaling} has also recently been observed to break down in numerous instances in systems with and without resetting \cite{HT09, NMV10, NT18, MeersonGaussian19, GM19, Jack20, BKLP20, MLMS21, MGM21, GIL21, GIL21b, Smith22OU}, 
 in which ``anomalous'' scalings were found: Namely, LDPs with exponents $\alpha$ and $\beta$ that are not both equal to 1.
 It is therefore appealing to search for anomalous scalings for reset processes too. 
The goal of this paper is to calculate the full distribution of the area under a broader class of processes that includes the RBM as a particular case: self-similar Gaussian processes (SGPs) with stochastic resetting (see below for a precise definition). This class also includes the reset fractional Brownian motion (rFBM) as a particular case, which was studied in \cite{MajumdarOshanin18, MM19}.

Here is the plan of the rest of the paper.
In section \ref{sec:mainResults}, we give a precise definition of the model and summarize our main findings.
In section \ref{sec:exact}, we derive an exact expression for the Fourier-Laplace transform of the distribution. The moderately-large-deviation behavior is then extracted from this expression in the long-time limit, uncovering a condensation transition that is subsequently characterized in detail. Then, the very-large-deviation regime is studied, in which yet another phase transition is found.
In section \ref{sec:cumulants} we argue that, despite the anomalous scaling, the cumulants all grow linearly in time at long times and, for the RBM, we derive a method for calculating them recursively.
In section \ref{sec:conclusion} we summarize our results and briefly discuss extensions.
Some technical details are relegated to the appendices.

\subsection{Model and summary of main results}
 \label{sec:mainResults}

Consider a Gaussian process $x(t)$ with zero mean and which is self-similar, i.e., $x(a\, t)\equiv a^H x(t)$ where $a$  is any constant and $H>0$ is a scaling exponent that characterizes the process. By $\equiv$, we mean that the trajectories of the two processes $x(a\,t)$ and $a^H\, x(t)$ have the same probability distribution over any duration. A Gaussian process
is completely characterized by its two-time correlation function 
$C\left(t_{1},t_{2}\right)=\left\langle x\left(t_{1}\right)x\left(t_{2}\right)\right\rangle $.
A consequence of the self-similarity is given by the scaling transformation 
\begin{equation}
C\left(a\,t_{1},a\,t_{2}\right)=a^{2H}\,C\left(t_{1},t_{2}\right)\,.
\label{ss_corr.1}
\end{equation}
There are several examples of self-similar Gaussian processes (SGPs). The most common example is the Brownian motion $B(t)$
which has $H=1/2$ and $C(t_1,t_2)= 2 D \min(t_1,t_2)$ where $D$ is the diffusion coefficient. A more general example is the so called fractional Brownian
motion (fBm) for which $C(t_1,t_2)= \mathcal{C} \left[t_1^{2H}+t_2^{2H}- |t_1-t_2|^{2H}\right]$ where $\mathcal{C}$ is a constant
and $0<H<1$ is called the Hurst exponent. The standard Brownian motion corresponds to $H=1/2$. Yet another
example of SGP is the so called random acceleration process, i.e., $x(t)= \int_0^t B(\tau) d\tau$
where $B(t)$ is a Brownian motion which starts at the origin, $B(0)=0$. This is called random acceleration, since $d^2x/dt^2= \sqrt{2D}\, \eta(t)$ where
$\eta(t)$ is a Gaussian white noise with zero mean and delta correlator $\langle \eta(t_1)\eta(t_2)\rangle = \delta(t_1-t_2)$.
For this random acceleration process, it is easy to see that $H=3/2$. A general SGP 
$x(t)$ has a scale $t^{H}$
and any moment $\langle x^p(t)\rangle = t^{p H} \mu_p$ where $\mu_p$ is a constant independent of $t$ for any $t$.

Now, consider a general SGP $x(t)$ (zero mean) with $H>0$ (that includes all these examples above as
special cases) and let 
\begin{equation}
A(t)= \int_0^t x(\tau)\, d\tau
\label{area.1}
\end{equation}
denote the area under such a process up to time $t$. Clearly, by linearity in Eq. (\ref{area.1}), it follows
that $A(t)$ is also a Gaussian process and in particular, its marginal distribution for fixed $t$ is
a Gaussian with zero mean and a variance $V(t)= \langle A^2(t)\rangle= \int_0^{t}\int_0^{t} C(t_1,t_2)\, dt_1\, dt_2$.
Writing $t_1= \tau_1 \, t$ and $t_2= \tau_2\, t$, and using the self-similar property in Eq. (\ref{ss_corr.1})
(choosing $a=t$), it follows that for any $t$
\begin{equation}
V(t)= \langle A^2(t)\rangle = 2\, c\, t^{2(H+1)}\, \quad {\rm where}\quad 2\,c= \int_0^1\int_0^1 C(u_1, u_2) du_1\, du_2\, .
\label{Vt.1}
\end{equation}
The constant $c$ and the index $H$ depend on the particular process. For example, for the simple Brownian motion, we have
$H=1/2$ and $c= D/3$. Hence, the marginal distribution $P_0(A|t)$ for any fixed $t$ is given by
\begin{equation}
P_0(A|t) = \frac{1}{\sqrt{4\, \pi\, c\, t^{2(H+1)}}}\, \exp\left[- \frac{A^2}{4\, c\, t^{2(H+1)}} \right]\, .
\label{P0At.1}
\end{equation} 
Note that we use the subscript $0$ to indicate that this is the free process without resetting---resetting will
be introduced shortly.

We now consider this general SGP with stochastic resetting to the origin at a constant rate $r$. This means that the PDF of the time interval $\tau$ between two successive resettings is simply $r\, e^{-r\, \tau}$. Our goal is to compute the marginal distribution $P_r(A|t)$ of the area $A(t)$ under the curve for a fixed $t$.

Here is a simple physical setting in which this observable is pertinent. Consider a ``physical" Brownian motion whose position and velocity evolve as
\be
dx/dt=v\,, \qquad  dv/dt= -\gamma v + \eta(t)\,,
\ee
where $\eta(t)$ is the usual white noise and $\gamma$ is the friction/damping coefficient. 
In the overdamped limit $t\gg 1/\gamma$,  the 
velocity itself becomes proportional to the noise,
$v(t)\approx (1/\gamma)\, \eta(t)$ and hence $x(t) \approx (1/\gamma)B(t)$
reduces, up to the damping constant $\gamma$ to the Wiener
process $B(t)=\int_0^t \eta(\tau)\, d\tau$.
However, in the underdamped limit $t \ll 1/\gamma$, the velocity $v(t) = B(t)$ and the position $x(t)$ is actually the area under a Wiener process $x(t)= \int_0^t B(\tau) d\tau$.
Now consider the situation where the velocity $v(t)$ of this `physical' Brownian motion is reset with a constant rate $r$ to its value 0. Then the area $A_i= \int_0^{t_i} v(\tau) d\tau$ during the i-th interval is precisely 
the physical displacement $\Delta x_i$ during the i-th interval. Hence $P_r(A|t)$ is precisely the position distribution of the
`physical' RBM at time t. 
Now for $t \ll 1/\gamma$ since $v(t)=B(t)$, our results for $P_r(A|t)$ will describe the position distribution of
a underdamped physical Brownian motion. In particular, our results for the large deviations at late times will be valid
when  $1/r\ll t \ll 1/\gamma$.

\medskip
 
\begin{figure}
\includegraphics[width=0.56\linewidth]{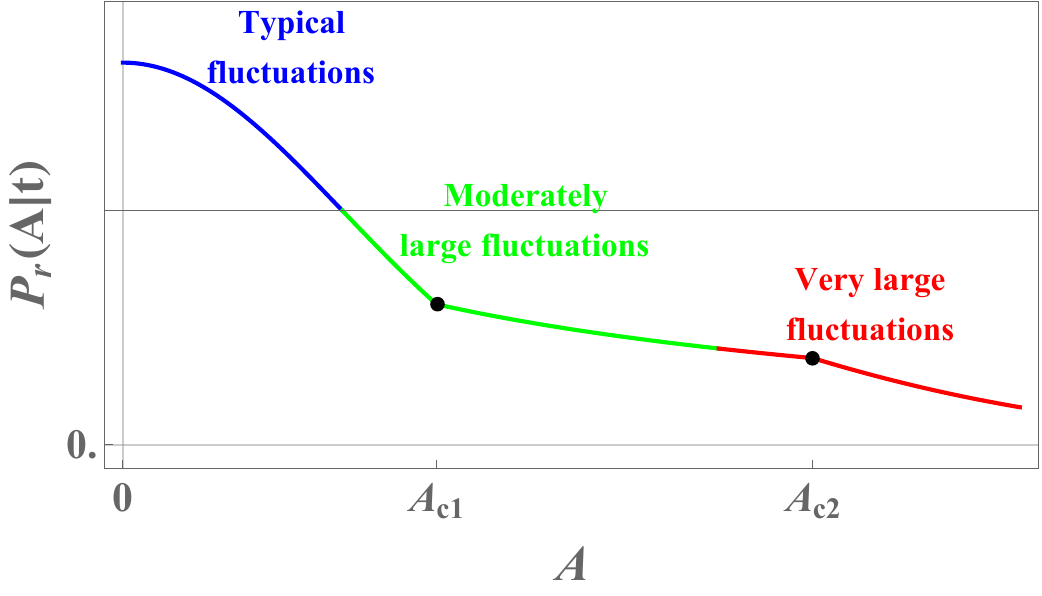}
\caption{Schematic plot of $P_r(A|t)$ vs $A$ at long times $t\gg1/r$. The three regimes in the figure correspond to the three cases in Eq.~\eqref{scalingRegimes}.
At $A = A_{c1} \sim t^{\beta}$ a first-order condensation transition occurs, separating between a homogeneous ($A<A_{c1}$) and a condensed ($A>A_{c1}$) phase.
At $A = A_{c2} \sim t^{(2H+3)/2}$ a second-order transition occurs, separating between a phase $A<A_{c2}$ in which dominant realizations reset $O(rt)$ times and a phase $A>A_{c2}$ in which the number of resetting events is $O(1)$.
The distribution is symmetric, $P_r(-A|t)=P_r(A|t)$, due to the mirror symmetry of the problem. Hence, only $A>0$ is plotted.}
\label{FigPrAtSchematic}
\end{figure}

Now, let us give a brief summary of our main results.
In the long time limit, $t \gg 1/r$, we identify three different regimes of $P_r(A|t)$:
when $A \sim \sqrt{t}$ (typical fluctuations), $A\sim t^{(2H+3)/(4H+4)}$ (large fluctuations)
and $A\sim t^{(2H+3)/2}$ (very large fluctuations).
The behavior of $P_r(A|t)$ can be summarized as
\be
\label{scalingRegimes}
P_{r}(A|t)\sim\begin{cases}
\exp\left[-\frac{r^{2H+1}}{4c\,\Gamma(2H+3)}\,\frac{A^{2}}{t}\right] & \text{for }|A|\sim O\left(\sqrt{t}\right)\text{ (typical fluctuations)},\\\\
\exp\left[-(rt)^{\alpha}\,\Phi\left(\sqrt{\frac{2}{c\,\Gamma(2H+3)}}\,r^{H+1-\beta}\,\frac{A}{t^{\beta}}\right)\right] & \text{for }|A|\sim O\left(t^{\beta}\right)\text{ (moderately large fluctuations)},\\\\
\exp\left[-rt\Psi\left(\frac{A}{\sqrt{cr}t^{\left(2H+3\right)/2}}\right)\right] & \text{for }|A|\sim O\left(t^{\left(2H+3\right)/2}\right)\text{ (very large fluctuations)},
\end{cases}
\ee
where $\Phi(y)$ has the leading-order asymptotic behaviors
\be
\label{PhiAsymptotics}
\Phi\left(y\right)=\begin{cases}
y^{2}/8 \,, & y\ll1\,,\\[3mm]
a_{0}\frac{\left(2H+3\right)}{\left(2H+2\right)}\left[\frac{\left(H+1\right)}{4a_{0}}y^{2}\right]^{1/\left(2H+3\right)}+\dots\,, & y\gg1\,,
\end{cases}
\ee
where $a_0= (\Gamma(2H+3))^{1/(2H+2)}$, and
\be
\label{Psidef}
\Psi\left(w\right)=\begin{cases}
\frac{2H+3}{2H+2}\left(\frac{H+1}{2}\right)^{1/\left(2H+3\right)}w^{2/\left(2H+3\right)}\,, & w<\sqrt{\frac{2}{H+1}}\,,\\[3mm]
\frac{1}{4}w^{2}+1\,, & w>\sqrt{\frac{2}{H+1}}\,,
\end{cases} 
\ee
see figure \ref{FigPrAtSchematic} for a schematic plot of $P_r(A|t)$ at long times. 
In both large-fluctuation regimes, these results constitute LDPs with exponents that differ from those of ``standard'' case. In the moderately-large-fluctuations regime they are given by
$\alpha=1/\left(2H+2\right)$ and $\beta=\left(2H+3\right)/\left(4H+4\right)$.
We calculate the rate functions $\Phi(y)$ and $\Psi(w)$ exactly: See Eqs.~\eqref{ratef.1} and \eqref{ratef.alternative} for two different (but equivalent) forms of $\Phi(y)$ (the equivalence is shown in Appendix \ref{app:PhiEquivalence}), and Eq.~\eqref{LDPPsi} for $\Psi(w)$.
We find that the behavior of $P_r(A|t)$ matches smoothly between the three regimes. This is seen from the 
asymptotic behavior given in the first line of Eq.~\eqref{PhiAsymptotics}
from which it follows that the first two regimes have a common regime of validity $\sqrt{t} \ll |A| \ll t^{\beta}$, and similarly from the behaviors given in the second line of Eq.~\eqref{PhiAsymptotics} and the first line of \eqref{Psidef} (which is valid, in particular, at $w\ll1$) that imply that the second and third regime are both valid at $t^{\beta} \ll |A| \ll t^{(2H+3)/2}$.

Remarkably, $\Phi(y)$ exhibits a first-order dynamical phase transition -- a discontinuity of its first derivative -- at a critical value $y=y_c$ which is given in \eqref{yc_final} below. In the subcritical regime $y<y_c$, $\Phi(y)$ is exactly parabolic, describing a Gaussian distribution of typical fluctuations of $A(t)$, and the system is in a ``homogeneous'' phase meaning that the realizations that dominate the contribution to $P_r(A|t)$ are those for which $A(\tau)$ grows (roughly) linearly in time, from time $\tau=0$ until time $\tau=t$.
In contrast, in the supercritical regime $y>y_c$, the system is in a ``condensed'' phase in which the dominant realizations are those for which $A(\tau)$ includes a temporally localized ``burst'', on top of the linear growth in time. This burst occurs at some intermediate time between 0 and $t$, and it corresponds to a single run of the process $x(\tau)$ in which no resetting occurs and under which a relatively large area is attained.

Moreover, the rate function $\Psi(w)$ that describes the very-large-fluctuations regime exhibits a second-order dynamical phase transition at the critical value $w=w_c=\sqrt{2/(H+1)}$ . This transition separates between a  regime $w<w_c$ in which the number of resetting events for the dominant realizations is of order $O(rt)$, and a  regime $w>w_c$ in which dominant realizations include a single run of the process that lasts for (nearly) the entire dynamics, so that the number of resetting events is $O(1)$.

Finally, we find that all of the cumulants of the distribution grow linearly with time at large $t$, i.e. $\langle A^n\rangle_c\approx c_n \, t\, .$, with coefficients $c_n$ that we calculate via a recursive relation, see Eq.~\eqref{cumulantsUpTo12} for the first 6 nonvanishing coefficients.
Remarkably, and in contrast to the ``usual'' case, we find no clear connection between the cumulants and the rate functions that describe the large-deviation regimes. Rather, the cumulants describe the corrections to the Gaussian behavior in the regime of typical fluctuations, $A(t) \sim \sqrt{t}$.

\section{The area under a self-similar Gaussian process with resetting: Large deviations and condensation}
\label{sec:exact}

\subsection{Exact Fourier-Laplace transform of $P_r(A|t)$ and asymptotics for long times and moderately large deviations}

In this section, we calculate the exact Fourier-Laplace transform of $P_r(A|t)$, and then extract its large-deviation behavior at long times.

We begin by noting that the Fourier transform of $P_0(A|t)$ in Eq.~\eqref{P0At.1} is simply
\begin{equation}
{\tilde P}_0(k|t)= \int_{-\infty}^{\infty} P_0(A|t)\, e^{i\, k\, A}\, dA= e^{- c\,t^{2(H+1)}\, k^2}\, .
\label{P0kt.1}
\end{equation} 
Also, for later purposes, let us define the Fourier-Laplace transform
\begin{equation}
{\tilde P}_0(k,s)= \int_0^{\infty} dt\, e^{-s\,t}\, {\tilde P}_0(k|t)= \int_0^{\infty} dt\, e^{-s\, t-  c\,t^{2(H+1)}\, k^2}\, .
\label{P0ks.1}
\end{equation}

\begin{figure}
\includegraphics[width=0.55\linewidth]{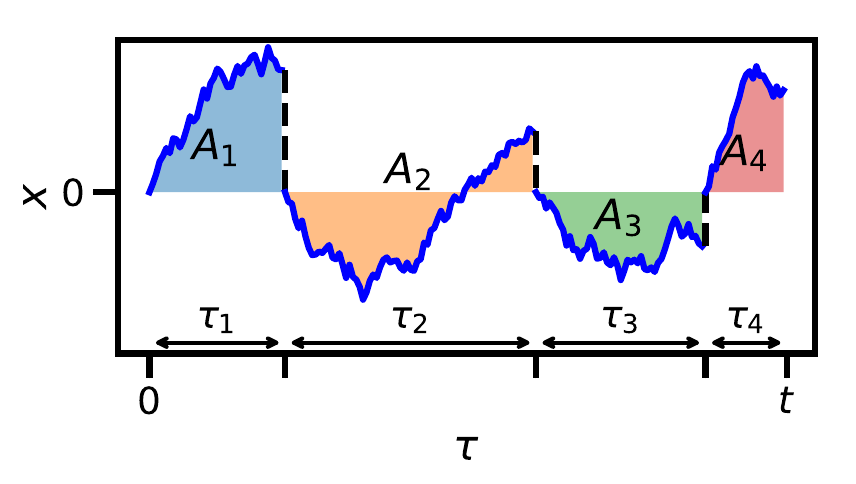}
\caption{Schematic plot of $x(\tau)$ with $n=4$ reset-free intervals.
When resetting occurs, corresponding the dashed lines in the figure, the position of the particle is set to the origin, $x=0$.
$\tau_i$ and $A_i$ are the duration and the area under $x(\tau)$, respectively, for the $i^{\text{th}}$ interval.
Area under the $\tau$ axis counts as negative.}
\label{FigSchematicRBM}
\end{figure}

To compute $P_r(A|t)$, let
$n = 1,2,\dots$ denote the number of time intervals between resetting events until time $t$, 
and $\vec{\tau}=\{\tau_1,\,\tau_2,\ldots,\, \tau_n\}$ denote the durations of these intervals,
so that the number of resettings until time $t$ is $n-1$.
Clearly, $\vec \tau$ and $n$ are both random variables.
Let $P_r(A, \vec \tau, n|t)$ denote the joint distribution of $A$, $\vec \tau$ and $n$ until time $t$.
Using the fact that after each resetting the process renews itself, this joint distribution reads
\begin{equation}
P_r(A,\vec \tau, n|t)= \int_{-\infty}^{\infty}dA_{1}\cdots\int_{-\infty}^{\infty}dA_{n}
\left[\prod_{i=1}^{n-1} r\, e^{-r\,\tau_i}\, P_0(A_i|\tau_i)\right]\, 
e^{-r\,\tau_n}\, P_0(A_n|\tau_n)\,
\delta\left(\sum_{i=1}^n \tau_i-t\right)\, \delta\left(\sum_{i=1}^n A_i-A\right)\, .
\label{pjoint.1}
\end{equation}
Here $A_i$ denotes the area under the $i^\text{th}$ run of the process (which is of duration $\tau_i$), between the resetting events $i-1$ and $i$, see Fig.~\ref{FigSchematicRBM} for an illustration.
Note that, unlike its predecessors, the weight of the last interval is $e^{-r\tau_n}$ (and not $r\, e^{-r \tau_n}$)
since the last interval is yet to be completed. 
Taking a Fourier transform with respect to $A$, a Laplace transform with respect to $t$ 
and integrating over $\vec \tau$, we get
\begin{equation}
\int_{-\infty}^{\infty}dA \, e^{i\, k\, A}\, \int_0^{\infty} dt\, e^{-s\, t}\, P_r(A,n|t)
= \frac{1}{r}\, \left[r \, {\tilde P}_0(k,r+s)\right]^n\, ,
\label{pjoint.2}
\end{equation}
where $P_r(A,n|t)$ is the joint distribution of $A(t)$ and $n$ at time $t$, and ${\tilde P}_0(k,s)$ is given in Eq. (\ref{P0ks.1}). Summing Eq.~\eqref{pjoint.2} over $n=1,2,\ldots$,
gives us the exact Fourier-Laplace transform of our desired marginal distribution
\begin{equation}
\int_{-\infty}^{\infty} dA \, e^{i\, k\, A}\, \int_0^{\infty} dt\, e^{-s\, t} \, P_r(A|t)= 
\frac{{\tilde P}_0(k, s+r)}{1-r\, {\tilde P}_0(k, s+r)}\, ,
\label{main_res.1}
\end{equation}
where we recall from Eq. (\ref{P0ks.1}) that
\begin{equation}
{\tilde{P}}_{0}(k,s+r)=\int_{0}^{\infty}dt\,\exp\left[-(s+r)\,t-c\,t^{2(H+1)}\,k^{2}\right]=\frac{1}{s+r}\,\int_{0}^{\infty}d\tau\,\exp\left[-\tau-\frac{c\,k^{2}}{(s+r)^{2H+2}}\,\tau^{2H+2}\right]\,.
\label{P0ksr.1}
\end{equation}
Note that the geometric series obtained when summing Eq.~\eqref{pjoint.2} over $n=1,2,\ldots$ always converges since $ r \tilde{P}_0(k, s+r) <1$ for any real $k$. This is easily seen by using $r/(r+s)<1$ while noticing that the integral over $\tau$ in Eq.~\eqref{P0ksr.1} is smaller than $1$ (since the integrand is smaller than $e^{-\tau}$).
Eq.~\eqref{main_res.1} was also derived in Ref. \cite{DH2019}
for RBM, but our derivation clearly shows that it is valid for
any stochastically reset process. Furthermore, in Ref. \cite{DH2019},
the large deviation behaviors of $P_r(A|t)$, using Eq.~\eqref{main_res.1},
were not analysed. Here we show below how Eq.~\eqref{main_res.1} can be successfully
used to extract the moderately large deviation behaviors.
Finally, inverting the Fourier and the Laplace transform in Eq. (\ref{main_res.1}) formally, we get
the main exact result
\begin{equation}
P_r(A|t)= 
\int_{-\infty}^{\infty} \frac{dk}{2\pi}\, e^{-i\, k\, A}\, \int_{\Gamma_0} \frac{ds}{2\pi i}\, e^{s\,t}\,
\frac{{\tilde P}_0(k, s+r)}{1-r\, {\tilde P}_0(k, s+r)}\, ,
\label{main_res.2}
\end{equation}
where $\Gamma_0$ is a Bromwich contour in the complex $s$ plane.
Upon substituting Eq. (\ref{P0ksr.1}) on the right hand side (rhs) of Eq. (\ref{main_res.2}), it turns
out to be convenient to rescale $k= (s+r)^{H+1}\, q$. This gives, after simple manipulation,
\begin{equation}
P_r(A|t)= \int_{-\infty}^{\infty} \frac{dq}{2\pi}\, f(q)\, \int_{\Gamma_0} \frac{ds}{2\pi i}\, e^{s\, t}\, (s+r)^{H+1}\,
\frac{ e^{-i\, (s+r)^{H+1}\, q\, A}}{s+r\, (1- f(q))}\, ,
\label{main_res.3}
\end{equation}
where the function $f(q)$ is given by
\begin{equation}
f(q)= \int_0^{\infty} d\tau\, e^{-\tau- c\, q^2\, \tau^{2H+2}}\, .
\label{fq_def}
\end{equation}
Note that the result in Eq. (\ref{main_res.3}) is exact at all times, since we haven't made any approximation
so far. 

Unfortunately, we can not evaluate the double integral on the rhs of Eq. (\ref{main_res.3}) exactly for any given $t$.
However, for large $t$, one can make progress as follows.
For large $t$, the dominant contribution in the integral over $s$ in Eq. (\ref{main_res.3}) comes from the vicinity of $s=0$.
Indeed, we will see soon that we will work in the scaling limit when both $t$ and $A$ are large (correspondingly
the conjugate variables $s$ and $q$ are small), with the ratio $A/t^{\beta}$ fixed, where the exponent $\beta$ will
be chosen appropriately.
Hence, to leading order for large $t$, one gets
\begin{equation}
P_r(A|t) \approx r^{H+1}\, \int_{-\infty}^{\infty} \frac{dq}{2\pi}\, f(q)\, e^{-i\, r^{H+1}\, q\, A}\,
\int_{\Gamma_0} \frac{ds}{2\pi i}\, \frac{e^{st}}{s+r (1-f(q))}\, .
\label{larget.1}
\end{equation}
Note that we did not perform a small $s$ expansion of $1/(s+ r(1-f(q))$ on the rhs of Eq. (\ref{main_res.3})
since $(1-f(q))$ is also small for small $q$. Since, our intention is to work in appropriate scaling limits,
we kept the denominator $s+r (1-f(q))$ as it is. With this approximation, the Bromwich integral
on the rhs of Eq. (\ref{larget.1}) can now be performed explicitly since it amounts to evaluating
the residue at the pole $s=- r(1-f(q))$ in the complex $s$ plane. Hence, Eq. (\ref{larget.1})
simplifies to
\begin{equation}
P_{r}(A|t)\approx r^{H+1}\,\int_{-\infty}^{\infty}\frac{dq}{2\pi}\,f(q)\,\exp\left[-i\,r^{H+1}\,q\,A-r\left(1-f(q)\right)\,t\right]\,.
\label{larget.2}
\end{equation}
Note that we have implicitly assumed that $1-f(q)$ is small for small $q$. In fact, from the
definition in Eq. (\ref{fq_def}),  expanding for small $q$, we get
\begin{equation}
f(q)= 1- c\, \Gamma(2H+3)\, q^2 + O(q^4)\, .
\label{small_q.1}
\end{equation}
Substituting $1-f(q)= c\, \Gamma(2H+3)\, q^2+O(q^4)$ in the term multiplying $t$ 
inside the exponential in Eq. (\ref{larget.2})
gives, up to the leading order $q^2$,
\begin{eqnarray}
P_r(A|t) &\approx & r^{H+1}\,\int_{-\infty}^{\infty}\frac{dq}{2\pi}\,f(q)\,\exp\left[-i\,r^{H+1}\,q\,A-r\,c\,\Gamma(2H+3)\,q^{2}\,t\right] \nonumber \\
&\approx &
r^{H+1}\,\int_{-\infty}^{\infty}\frac{dq}{2\pi}\,\int_{0}^{\infty}d\tau\,\exp\left[-\tau-c\,q^{2}\,\tau^{2H+2}-i\,r^{H+1}\,q\,A-r\,c\,\Gamma(2H+3)\,q^{2}\,t\right]\,,\label{larget.3}
\end{eqnarray}
where, in going from the first to the second line, we used the integral representation of $f(q)$ in Eq. (\ref{fq_def}).
To make different terms inside the exponential in Eq. (\ref{larget.3}) of the same order, we next make the following
rescalings
\begin{equation}
\tau= (rt)^{\alpha}\, u ; \quad \,\,  A= b\, t^{\beta}\,  y\, ;  \quad \,\,  q = (rt)^{-\gamma}\, {\tilde q}
\label{rescale.1}
\end{equation}
Our eventual goal is to evaluate the integral over $\tau$ by a saddle point method.
In order that all four terms inside the exponential in Eq. (\ref{larget.3}) are of the same order, 
it is easy to check that we must have:
(i) $\alpha\left(2H+2\right)-2\gamma$,
(ii)  $\beta-\gamma=\alpha$ and 
(iii) $1-2\gamma=\alpha$. Solving these relations give
us the unique choice
\begin{equation}
\alpha= \frac{1}{2H+2}\, ; \quad  \beta= \frac{2H+3}{4H+4}\,; \quad {\rm and}\,\,\, \gamma= \frac{2H+1}{4H+4}\, .
\label{rescale.2}
\end{equation}
Furthermore, Eq. (\ref{larget.3}) then reduces to
\begin{equation}
P_r(A|t)\approx r^{H+1}\, (rt)^{(1-2H)/(4H+4)}\, \int_{-\infty}^{\infty} \frac{d \tilde{q}}{2\pi}\, \int_0^{\infty}
du\, \exp\left[ - (rt)^{\alpha}\,\left( u+ 
c {\tilde q}^2\, u^{2H+2}\,+ c\, \Gamma(2H+3)\, {\tilde q}^2 + i\, b\, r^{H+1-\beta}\, {\tilde q}\, y\right)\right]\, ,
\label{larget.4}
\end{equation}
with $\alpha$ and $\beta$ given explicitly in Eq. (\ref{rescale.2}). Note that the scale factor $b$, for the moment, is free
and we can choose it at our convenience. Now, we can first perform the integral over ${\tilde q}$ exactly since
it is just a Gaussian integral. Using the identity
\begin{equation}
\int_{-\infty}^{\infty} \frac{d \tilde{q}}{2\pi} e^{-c_2 \tilde{q}^2 -i\, c_1\, \tilde{q}}= 
\frac{1}{\sqrt{4\,\pi\, c_2}}\, e^{-c_1^2/{4 c_2}}\, ; \quad {\rm for}\,\, c_2>0\, ,
\label{gaussian_int.1}
\end{equation}
we then get (ignoring pre-exponential factors)
\begin{equation}
 P_r(A|t)\sim \int_0^{\infty} du \, \exp\left[-(rt)^{\alpha}\, \left(u + \frac{b^2\, r^{2(H+1-\beta)}\, y^2}{4\, 
c\, (u^{2H+2}+\Gamma(2H+3))} \right)\right]\, ,
\label{larget.5}
\end{equation}
where we recall that $A= b\, t^{\beta}\, y$.
We can further simplify the integral by rescaling $u\to (\Gamma(2H+3))^{1/(2H+2)}\, u$ and
by choosing $b= \sqrt{c\,\Gamma(2H+3)/2}\, r^{-(H+1-\beta)}$. This gives
\begin{equation}
P_r(A|t)\sim \int_0^{\infty} du \, \exp\left[-(rt)^{\alpha}\, 
\left(a_0\, u+ \frac{y^2}{8(u^{2H+2}+1)}\right)\right]\, ; \quad {\rm where} \quad a_0= (\Gamma(2H+3))^{1/(2H+2)}\, .
\label{larget.6}
\end{equation}
Finally, evaluating the integral for large $t$ by the saddle point method, we arrive at our final result
\begin{equation}
P_r(A|t) \sim \exp\left[- (rt)^{\alpha}\, \Phi\left(y= \sqrt{\frac{2}{c\, \Gamma(2H+3)}}\, r^{H+1-\beta}\, 
\frac{A}{t^{\beta}}\right)\right]\, ,
\label{larget.7}
\end{equation}
where the rate function $\Phi(y)$ is given by     
\begin{equation}
\Phi(y)=\min_{u\in\left[0,\infty\right)}\left[a_{0}\,u+\frac{y^{2}}{8(u^{2H+2}+1)}\right]\,.
\label{ratef.1}
\end{equation} 
where we recall that the constant $c$ is defined in Eq. (\ref{Vt.1}) characterizing the process. The
exponents $\alpha$, $\beta$ and the constant $a_0$ are recalled as
\begin{equation}
\alpha= \frac{1}{2H+2}\, ; \quad \beta= \frac{2H+3}{4H+4}\,; \quad a_0=(\Gamma(2H+3))^{1/(2H+2)}\, .
\label{exp_const.1}
\end{equation} 
As an example, for the simple Brownian motion, using $H=1/2$ and $c=D/3$, our result in Eq. (\ref{larget.7})
predicts $\alpha=1/3$, $\beta=2/3$, $a_0= 6^{1/3}$ and
\begin{equation}
P_r(A|t)\Big|_{\rm BM}\sim \exp\left[- (rt)^{1/3}\, \Phi_{\rm BM}\left(y= \frac{r^{5/6}\, A}{\sqrt{D}\, t^{2/3}}\right)\right]\, ,
\label{BM.1}
\end{equation}
with the Brownian rate function given exactly by
\begin{equation}
\Phi_{\rm BM}(y)= \min_{u\in\left[0,\infty\right)} \left[ 6^{1/3}\, u + \frac{y^2}{8(u^{3}+1)}\right]\, .
\label{BMrf.1}
\end{equation}
As shown below, $\Phi(y) = y^2 / 8$ for subcritical $y$'s, describing a Gaussian distribution
\be
\label{PrAtGaussian}
P_{r}(A|t)\sim\exp\left[-\frac{r^{2H+1}}{4c\,\Gamma(2H+3)}\,\frac{A^{2}}{t}\right]
\ee
with a variance that grows linearly in time. For Brownian motion, $H=1/2$ the result \eqref{PrAtGaussian} agrees with that of \cite{DH2019}.

\subsection{The analysis of the critical point $y=y_c$ of the rate function $\Phi(y)$}

Due to the exact mirror symmetry of the problem, $P_{r}\left(A|t\right)=P_{r}\left(-A|t\right)$ and, as a result, $\Phi\left(y\right)=\Phi\left(-y\right)$. Therefore, for convenience we assume $y>0$ in this subsection.

The large deviation behavior of $P_r(A|t)$ for large $t$ and moderately large $A$ is described
in Eq.~(\ref{larget.7}) with the rate function $\Phi(y)$ given in Eq. (\ref{ratef.1}).
In this section, we will see that for any $H>0$, the rate function $\Phi(y)$ has
a singularity at $y_c$ where its first derivative $\Phi'(y)$ is discontinuous.
Physically, this point $y=y_c$ signals the onset of a condensed phase with a single
condensate. 
By `condensation' we mean that one of the terms in
the sum $A= \sum_i A_i$ is macroscopic, i.e., $A_i = O(A)$ for some $i$, which, as we show Section \ref{sec:condensedPhase} below, is indeed the case in the supercritical regime.
The rate function $\Phi(y)$ acts like an effective free energy and
at $y=y_c$ the transition is of first-order. We will see below that the mathematical
mechanism behind this first-order phase transition is exactly like in
standard thermodynamic phase transitions, once we interpret $\Phi(y)$ as an effective
free energy. Such a first-order phase transition characterizing a condensation
transition has been found recently in a number of works in different contexts in physics~\cite{GM19,MLMS21,
MGM21,GIL21, GIL21b,  Smith22OU}, 
as well as in the probability literature~\cite{BKLP20}.

\begin{figure}
\includegraphics[width=0.55\linewidth]{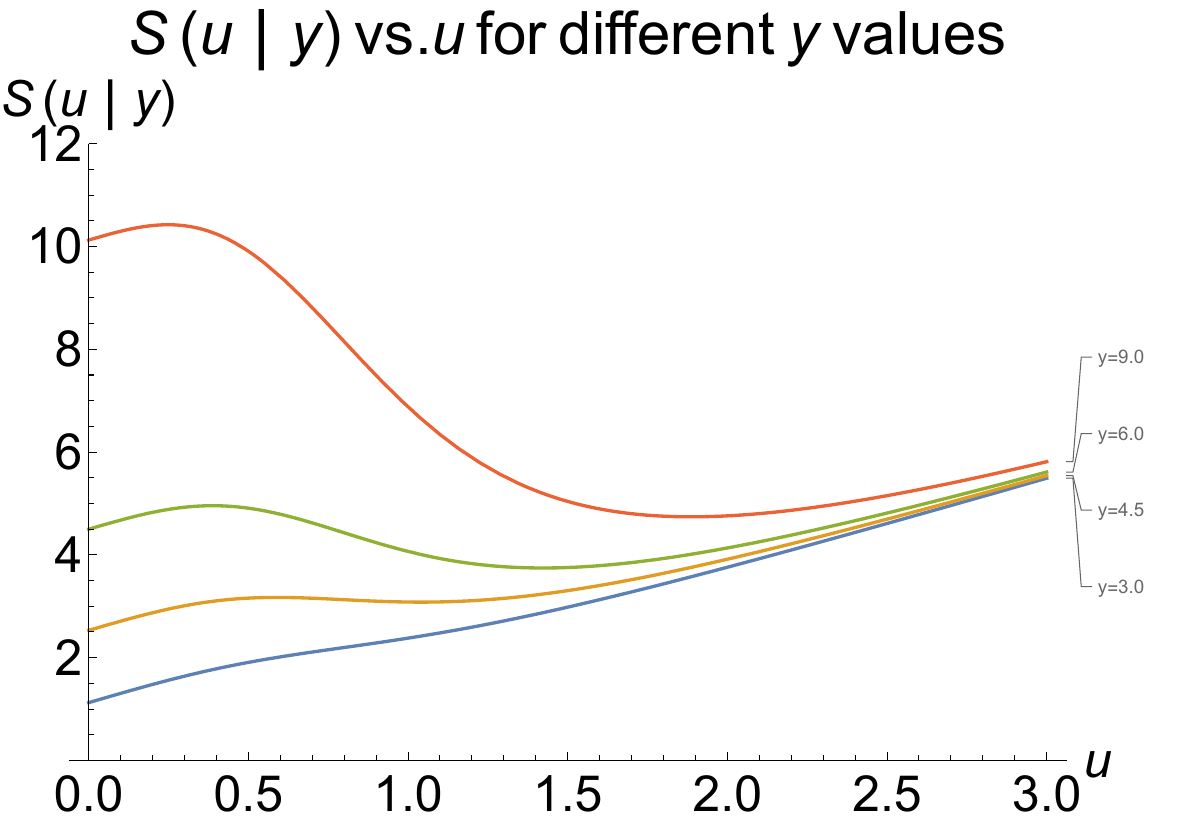}
\caption{Plot of $S(u|y)$ vs $u\ge 0$ for $H=1/2$ and different values of $y$.
For $y<y_1= 2 \times 3^{2/3}= 4.16017\dots$, the function $S(u|y)$ has a single minimum at $u=0$. When $y$ exceeds $y_1$, it develops  an additional, nonzero local minimum, which becomes the global minimum at $y > y_c= 2^{4/3}\, 3^{2/3}=5.24148 \dots$.
This is exactly the mechanism of a first-order phase transition once one interprets $\Phi(y)$ as an effective free energy.}
\label{su_BM_fig.1}
\end{figure}

To proceed, we consider $\Phi(y)$ in Eq. (\ref{ratef.1}) and write it as
\begin{equation}
\Phi(y)=\min_{u\in\left[0,\infty\right)}\left[S(u|y)\right]\,;\quad{\rm where}\quad S(u|y)=a_{0}\,u+\frac{y^{2}}{8(u^{2H+2}+1)}\,\quad{\rm with}\quad a_{0}=(\Gamma(2H+3))^{1/(2H+2)}\,.
\label{Suy.1}
\end{equation}
It is instructive to plot the function $S(u|y)$ vs. $u\ge 0$ for different values
of $y$ (see Fig. \ref{su_BM_fig.1} for the Brownian case $H=1/2$). It turns
out that as long as $y<y_1$, the function $S(u|y)$ has a single minimum
at $u=0$. When $y>y_1$, the function develops a new pair
of local maximum and local minimum respectively at $u_{-}(y)$ and $u_+(y)$.
For $y_1<u<y_c$, the minimum at $u=0$ remains the global minimum, i.e.,
$S(0|y)< S(u_+(y)|y)$. However, as $y$ exceeds a critical value $y_c>y_1$, 
the minimum at $u_+$ takes over as the global minimum, i.e., $S(u_+(y)|y)<S(0|y)$.
This competition between the two local minimum is a hallmark of a first-order
phase transition in thermodynamics. Indeed at $y=y_c$, the effective free energy
$\Phi(y)$ develops a first-order singularity, i.e., the first derivative
$\Phi'(y)$ is discontinuous at $y=y_c$. Below, we will compute $y_1$ and $y_c$
explicitly.

\begin{figure}
\includegraphics[width=0.52\linewidth]{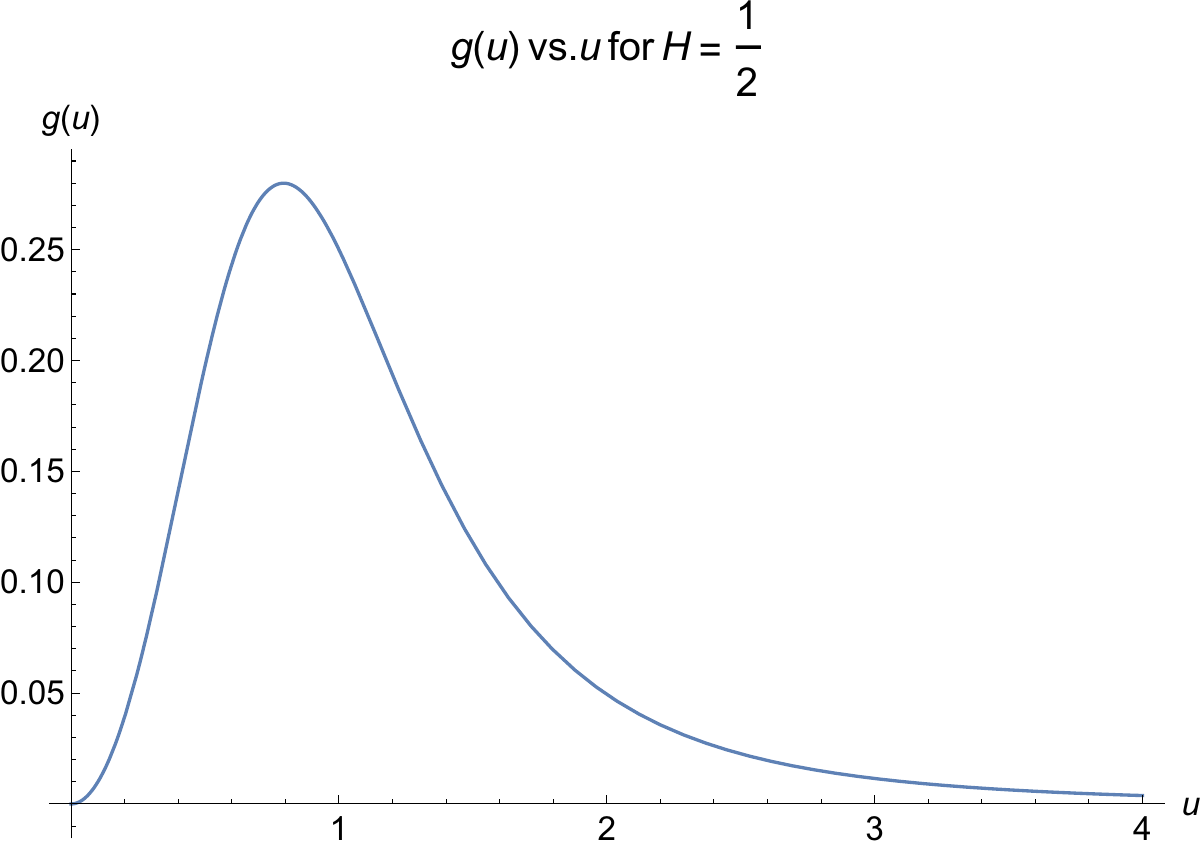}
\caption{Plot of $g(u)= u^2/(1+u^3)^2$ vs. $u$ for $H=1/2$. The curve has a single maximum at $u_m=2^{-1/3}$.}
\label{gu_BM_fig.2}
\end{figure}

To minimize $S(u|y)$ in Eq. (\ref{Suy.1}) for fixed $y$, we set $S'(u|y)=0$. This gives
the equation for the saddle point
\begin{equation}
\frac{8 a_0}{(2H+2)\, y^2}= \frac{u^{2H+1}}{\left(1+ u^{2H+2}\right)^2}\equiv g(u)\, .
\label{gu_def}
\end{equation}
The solution $u^*>0$ of this equation, if it exists, will provide a minimum.
To see if there is such a solution for a given $y$, let us first plot $g(u)$ vs $u$
(see Fig. \ref{gu_BM_fig.2} for the case $H=1/2$). This curve has a single maximum
at $u=u_m= [(2H+1)/(2H+3)]^{1/(2H+2)}$ with the maximum value $g(u_m)=
u_m^{2H+1}/(1+ u_m^{2H+2})^2$. From the saddle point equation (\ref{gu_def}),
it follows that if the left hand side (lhs) $8\, a_0\, y^{-2}/(2H+2)$ exceeds $g(u_m)$,
there is no solution $u^*$ to this saddle point. Hence, a nontrivial saddle
point $u^*$ exists only when $y>y_1$ where
\begin{equation}
\frac{8\, a_0}{(2H+2)\, y_1^2}= g(u_m)= 
\frac{u_m^{2H+1}}{\left(1+ u_m^{2H+2}\right)^2}\, ; \quad {\rm where}\quad u_m= \left(\frac{2H+1}{2H+3}\right)^{1/(2H+2)}\, .
\label{y1.1}
\end{equation}
This gives $y_1$ explicitly for any $H$. For instance, for $H=1/2$, it gives
\begin{equation}
y_1= 2 \times 3^{2/3}=4.16017\ldots
\label{y1_BM}
\end{equation}

When $y$ exceeds $y_1$, the saddle point equation (\ref{gu_def}) has two solutions
$u_{-}(y)$ and $u_+(y)$ with $u_{-}(y)<u_m< u_{+}(y)$. It is easy to check that $u=u_{-}(y)$ 
corresponds
to the local maximum of $S(u|y)$ at $u=u_{-}(y)$, while $u_+(y)$ corresponds to
an additional local minumum of $S(u|y)$ (see Fig. \ref{su_BM_fig.1} for $H=1/2$).
Thus for $y>y_1$, we have two local minima of $S(u|y)$: one at $u=0$ and one at $u=u_+(y)$.
Hence, we need to now compare the value of the action at the two minima, namely
$S(u=0|y)$ and $ S(u=u_+(y)|y)$ to see which one is the global minimum. Now, from 
Eq. (\ref{Suy.1}) we get
\begin{equation}
S(0|y)= \frac{y^2}{8}\, .
\label{S0.1}
\end{equation}
We anticipate (see Fig. \ref{su_BM_fig.1}) that for $y_1<y<y_c$, the minimum
at $u=0$ will be the global minimum, while for $y>y_c$, it will be taken over
by the minimum at $u_+(y)$. Hence the critical value $y_c$ is determined from the
condition
\begin{equation}
S(0|y_c)= S(u=u_+(y_c)|y_c)\, ; \quad {\rm implying}\quad \frac{y_c^2}{8}= a_0\, u_+(y_c) + 
\frac{y_c^2}{8 
\left(1+ u_+(y_c)^{2H+2}\right)}\, .
\label{yc.1}
\end{equation}
Simplifying, we get
\begin{equation}
\frac{y_c^2}{8 a_0}= \frac{1+ u_+(y_c)^{2H+2}}{u_+(y_c)^{(2H+1)}}\, .
\label{yc.2}
\end{equation}
On the other hand, putting $u=u_+(y_c)$ in the saddle point equation (\ref{gu_def}) gives
\begin{equation}
\frac{y_c^2}{8 a_0}= \frac{\left(1+ u_+(y_c)^{2H+2}\right)^2}{(2H+2)\, u_+(y_c)^{2H+1}}\, .
\label{yc.3}
\end{equation}
Equating the rhs of Eqs. (\ref{yc.2}) and (\ref{yc.3}) gives that at $y=y_c$
\begin{equation}
u_+ (y_c)= (2H+1)^{1/(2H+2)}\, .
\label{uplus_crit.1}
\end{equation}
Plugging this value on the rhs of Eq. (\ref{yc.3}) then gives us the value of $y_c$
\begin{equation}
y_c= \frac{4\, \sqrt{(H+1)\, a_0}}{(2H+1)^{(2H+1)/(4H+4)}}= 4\, \sqrt{H+1}\, \left[
\frac{\Gamma(2H+3)}{(2H+1)^{2H+1}}\right]^{1/(4H+4)}\, ,
\label{yc_final}
\end{equation}
where we used the explicit expression for $a_0= \left(\Gamma(2H+3)\right)^{1/(2H+2)}$.
For example, for the simple Browian motion case $H=1/2$, this gives
\begin{equation}
y_c\Big|_{\rm BM}= 2^{4/3}\, 3^{2/3}= 5.24148\ldots
\label{yc_BM.2}
\end{equation}

\begin{figure}
\includegraphics[width=0.52\linewidth]{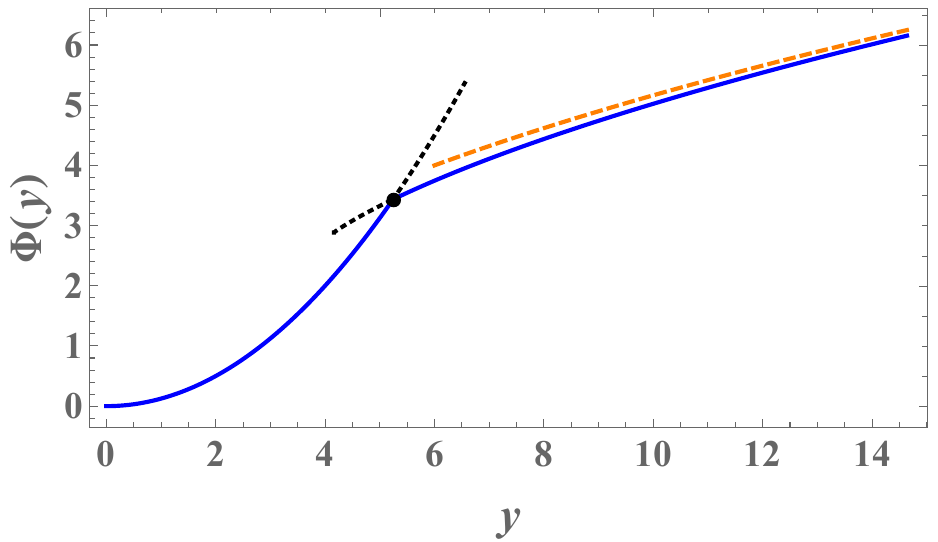}
\caption{Plot of $\Phi(y)$ vs. $y$ (solid line) for the Brownian case $H=1/2$.
The rate function has a singularity at $y=y_c=5.24148\ldots$ corresponding to the fat dot on the figure. For $y<y_c$, $\Phi(y)=y^2/8$, while it changes to $\Phi(y)=\chi(y)$ for $y>y_c$. 
At $y=y_c$, the first derivative of $\Phi(y)$ jumps, corresponding to a first-order dynamical phase transition.
The dotted lines are the continuations of the two non-optimal branches: $\chi(y)$ at $y_1 < y < y_c$ and $y^2/8$ at $y>y_c$.
The dashed line is the leading-order large-$y$ asymptotic behavior, given by the first term on the r.h.s. of Eq.~\eqref{PhiLargey}.}
\label{Phi_BM_with_chi}
\end{figure}

Hence, summarizing, the rate function $\Phi(y)$ in Eq. (\ref{ratef.1}) can be written as
\begin{eqnarray}
\label{ratef.2}
\Phi(y)=\begin{cases}
S(0|y)=\frac{y^{2}}{8}\,, & {\rm for}\quad y<y_{c}\,,\\[3mm]
S(u_{+}(y)|y)=\chi(y)\,, & {\rm for}\quad y>y_{c}\,,
\end{cases}\end{eqnarray}
where the function $\chi(y)$, for $y>y_c$, can be expressed parametrically as a function of $y$ by eliminating
$u$ from the pair of equations: $\chi= S(u|y)$ with $S(u|y)$ given in Eq. (\ref{Suy.1})
and the saddle point equation (\ref{gu_def}) connecting $y^2$ with $u$. More precisely, we
can write
\begin{eqnarray}
\chi &= & a_0\, \left[u+ \frac{(1+u^{2H+2})}{(2H+2)\, u^{2H+1}}\right]\,, \label{chi.1} \\
y^2 &= & \frac{4\,a_{0}}{H+1}\,\frac{\left(1+u^{2H+2}\right)^{2}}{u^{2H+1}}\,. \label{chi.2}
\end{eqnarray}
For example, for the simple Brownian motion where $H=1/2$ and $a_0= 6^{1/3}$
and $y_c= 2^{4/3} 3^{2/3}$, these pair
of equations for $y>y_c$ simplify to 
\begin{eqnarray}
\chi &= & 6^{1/3}\, \left[u+ \frac{(1+u^3)}{3\, u^{2}}\right]\,, \label{chi_BM.1} \\
y &= & \frac{2^{5/3}}{3^{1/3}}\, \frac{(1+u^{3})}{u}\,. \label{chi_BM.2}
\end{eqnarray}
For the Brownian case $H=1/2$, the full rate function $\Phi(y)$ in Eq. (\ref{ratef.2}) is plotted in Fig.~\ref{Phi_BM_with_chi}, where in the supercritical regime it is given by a parametric plot of $\chi(y)$ vs $y$ from Eqs. (\ref{chi_BM.1}) and (\ref{chi_BM.2}).
In the limit $y \gg 1$, we approximately solve \eqref{chi.2} for $u$ to get 
\be
u \simeq \left[\frac{\left(H+1\right)}{4a_{0}}y^{2}\right]^{1/\left(2H+3\right)}-\frac{2}{\left(2H+3\right)\left[\frac{\left(H+1\right)}{4a_{0}}y^{2}\right]^{\left(2H+1\right)/\left(2H+3\right)}} \, .
\ee
Using this in \eqref{chi.1} we find the asymptotic behavior
\be
\label{PhiLargey}
\Phi(y \gg 1) \simeq  a_{0}\frac{\left(2H+3\right)}{\left(2H+2\right)}\left[\frac{\left(H+1\right)}{4a_{0}}y^2\right]^{1/\left(2H+3\right)}-\frac{a_{0}}{\left(2H+2\right)}\left[\frac{\left(H+1\right)}{4a_{0}}y^2\right]^{-\left(2H+1\right)/\left(2H+3\right)} \, .
\ee
Plugging the leading-order term of \eqref{PhiLargey} into \eqref{larget.7}, we find that the tail of the moderately large fluctuations regime is given by a stretched exponential,
\be
\label{JointRegime}
P_{r}\left(A|t\right)\sim \exp\left\{ -r\,\frac{2H+3}{2H+2}\left[\frac{\left(H+1\right)A^{2}}{2cr}\right]^{1/\left(2H+3\right)}\right\} \, .
\ee
The leading-order asymptotic behaviors of $\Phi(y)$ are conveniently summarized as
\be
\label{PhiAsymptotics2}
\Phi\left(y\right)=\begin{cases}
y^{2}/8, & y\ll1\,,\\[3mm]
a_{0}\frac{\left(2H+3\right)}{\left(2H+2\right)}\left[\frac{\left(H+1\right)}{4a_{0}}y^{2}\right]^{1/\left(2H+3\right)}+\dots\,, & y\gg1\,.
\end{cases}
\ee

\subsection{Analysis of the condensed phase $A(t) > A_{c1}$}
\label{sec:condensedPhase}

In this subsection, we characterize the condensed phase $(y>y_c)$ of the moderately-large-fluctuations regime, $A(t) \sim t^\beta$ (where we recall that the exponent $\beta=(2H+3)/(4H+4)$), in some detail.
In analogy with condensation transitions found recently in several other systems \cite{GM19, MGM21, MLMS21, GIL21, GIL21b, MGM21, Smith22OU}, one can anticipate certain properties of the two phases found above.
We expect the subcritical phase, $A(t) / t^\beta < y_c$, to be ``homogeneous'', i.e., that for the realizations that contribute most to $P_r(A|t)$, the area grows homogeneously throughout the dynamics,
\be
\label{Athomogeneous}
A\left(\tau\right) \simeq \tau A\left(t\right)/t \, .
\ee
In contrast, we expect the supercritical phase, $A(t) / t^\beta > y_c$, to be ``condensed'': We anticipate dominant realizations to have a temporally localized ``burst'', so Eq.~\eqref{Athomogeneous} will be replaced by
\be
A\left(\tau\right)=A_{c}\theta\left(\tau-\tau_{c}\right)+\tau\left[A\left(t\right)-A_{c}\right]/t
\ee
where $A_c$ is the area attained at this localized burst (and is expected to be very large, of order $A(t)$), $\tau_c$ is its occurence time, and $\theta(\dots)$ is the Heaviside function.
We therefore expect the area under one of the runs to be very large. All of the runs are on equal footing, however, in the analysis below it is convenient to analyze the case in which the condensation occurs during the first run, and then to extend the result by using the symmetry between exchanging  different runs.
We now show that these arguments are indeed correct, by reproducing Eq.~\eqref{larget.7} using a different method. In doing so, we obtain another, equivalent representation of the rate function $\Phi(y)$.

As we will see below, it is useful to first calculate the distribution $P_{r}\left(A_{1}\right)$ of the area $A_1$ under the first run. It is given (exactly) by
\bea
\label{PrA1Exact}
P_{r}\left(A_{1}\right)&=&e^{-rt}P_{0}\left(A_{1},t\right)+r\int_{0}^{t}d\tau_{1}e^{-r\tau_{1}}P_{0}\left(A_{1},\tau_{1}\right) \nn\\
&=&\frac{1}{\sqrt{4\,\pi\,c\,t^{2(H+1)}}}e^{-\mathcal{S}\left(t\right)}+\int_{0}^{t}\frac{r}{\sqrt{4\,\pi\,c\,\tau_{1}^{2(H+1)}}}\,e^{-\mathcal{S}\left(\tau_{1}\right)}\,d\tau_{1},\quad\mathcal{S}\left(\tau_{1}\right)\equiv r\tau_{1}+\frac{A_{1}^{2}}{4\,c\,\tau_{1}^{2(H+1)}}\,,
\eea
where we used Eq.~\eqref{P0At.1}.
The first term in \eqref{PrA1Exact} corresponds to the case in which no resetting events occur throughout the dynamics. 
The second term corresponds to the case in which at least one resetting event occurs, with $\tau_1$ denoting the first resetting time (using the same conventions as in Eq.~\eqref{pjoint.1}).
Let us focus on large deviations of $A_1$ in the long-time limit $rt\gg1$.
In the large-$A_1$ limit, the integral over $\tau_1$ in \eqref{PrA1Exact} is well approximated by the saddle-point approximation. The integrand can be approximated simply by $e^{-\mathcal{S}\left(\tau_{1}\right)}$, with the factor $\frac{r}{\sqrt{4\,\pi\,c\,\tau_{1}^{2(H+1)}}}$ only contributing to the subleading prefactor.
The saddle-point equation thus becomes quite simply, $\mathcal{S}'\left(\tau_{1}\right)=0$, which is immediately solved,
\be
\label{tau1star}
\tau_1^* =\left(\frac{\left(H+1\right)A_{1}^{2}}{2cr}\right)^{1/\left(2H+3\right)} \, .
\ee
Now, we must separate the analysis into two cases. If $\tau_1^* < t$, then 
plugging it back into the integrand in \eqref{PrA1Exact}, we get a stretched-exponential tail
\be
\label{PrA1NearTail}
P_r\left(A_{1}\right)\sim e^{-\mathcal{S}\left(\tau_{1}^{*}\right)}=\exp\left[-\frac{2H+3}{2H+2}r^{\left(2H+2\right)/\left(2H+3\right)}\left(\frac{H+1}{2c}\right)^{1/\left(2H+3\right)}A_{1}^{2/\left(2H+3\right)}\right]\,,
\ee
the first term in \eqref{PrA1Exact} being negligible.
For the result \eqref{PrA1NearTail} to be valid, one must consider $A_1$ sufficiently large so that the absolute value of the expression inside the exponent is much larger than 1.
The result \eqref{PrA1NearTail} is in agreement with the tail of the exact result obtained in \cite{Meerson19} for $H=1/2$ (Brownian motion) in the context of a ``mortal'' Brownian particle.
For later reference, we note that for $A_1 / \sqrt{cr}t^{\left(2H+3\right)/2} \ll 1$, $\tau_1^* \ll t$.
The other case, $\tau_1^* > t$, is discussed below in subsection \ref{sec:veryLargeDeviations}.
A similar saddle point analysis was also used to derive
the position distribution of an RBM at late times in Ref.~\cite{MSS15a}, as discussed
below.

\medskip
We now turn back to the analysis of the distribution $P_{r}\left(A|t\right)$ of the total area.
Let us write the exact desired PDF as the sum of two terms,
\be
\label{PrAtSum}
P_{r}\left(A|t\right)=e^{-rt}P_{0}\left(A|t\right)+\int_{0}^{t}re^{-r\tau_{1}}d\tau_{1}\int_{-\infty}^{\infty}P_{0}\left(A_{1}|\tau_{1}\right)P_{r}\left(A-A_{1}|t-\tau_{1}\right)dA_{1} \, .
\ee
The first term corresponds to the case in which no resetting events occur throughout the dynamics. The second term corresponds to the case in which at least one resetting event occurs, and $\tau_1$ and $A_1$ are the first resetting time and the area under the first run of the dynamics respectively (using the same conventions as in Eq.~\eqref{pjoint.1}).
Let us focus on long times, $rt\gg1$, and moderately large deviations of $A(t) \sim t^{\beta}$, in this regime, the first term in \eqref{PrAtSum} is negligible, because $e^{-rt} \ll P_{r}\left(A|t\right)$ as can be seen from the scaling $ -\ln P_{r}\left(A|t\right) \sim t^{\alpha}$ in \eqref{larget.7}.
Anticipating the existence of a condensed phase in which a macroscopic fraction of the area is attained during one of the runs, we look for a dominant contribution to the integral in the second term in \eqref{PrAtSum} that involves a large area under the first run, $A_1 \sim A$.
We thus aim to evaluate the double integral in \eqref{PrAtSum} via a saddle-point approximation on both $A_1$ and $\tau_1$.
A simplification arises for $A \ll t^{(2H+3)/2}$ (this includes the moderately large fluctuations regime):
Here, from Eq.~\eqref{tau1star} we see that the ``optimal'' $\tau_1$ is much smaller than $t$ so that $t-\tau_1$ can be replaced by $t$ in \eqref{PrAtSum}.
We thus arrive at
\be
\label{PrAtapprox}
P_{r}\left(A|t\right)\sim\int_{-\infty}^{\infty}dA_{1}P_{r}\left(A-A_{1}|t\right)\int_{0}^{t}d\tau_{1}\,re^{-r\tau_{1}}P_{0}\left(A_{1}|\tau_{1}\right)\,.
\ee
We identify that the integral over $\tau_1$ in \eqref{PrAtapprox} coincides exactly with the integral over $\tau_1$ in \eqref{PrA1Exact}. At $A_1 \sim t^{\beta}$ this integral yields $P_r(A_1)$, as we showed above. Therefore, Eq.~\eqref{PrAtapprox} becomes, at $A \sim t^{\beta}$,
\be
\label{PrAtIntegralA1}
P_{r}\left(A|t\right)\sim\int_{-\infty}^{\infty}P_{r}\left(A_{1}\right)P_{r}\left(A-A_{1}|t\right)dA_{1}\,.
\ee

We further assume (and check this assumption aposteriori in Appendix \ref{app:singleCondensate}) that the saddle point that we are after is in the regime  in which the term $P_{r}\left(A-A_{1}|t\right)$ is approximated well by the Gaussian distribution \eqref{PrAtGaussian}. 
Plugging Eqs.~\eqref{PrA1NearTail} and \eqref{PrAtGaussian} into \eqref{PrAtIntegralA1}, we obtain
\be
\label{PrAtIntegralOverA1}
P_{r}\left(A|t\right)\sim\int_{-\infty}^{\infty}\exp\left[-\frac{2H+3}{2H+2}r^{\left(2H+2\right)/\left(2H+3\right)}\left(\frac{H+1}{2c}\right)^{1/\left(2H+3\right)}A_{1}^{2/\left(2H+3\right)}-\frac{r^{2H+1}}{4c\,\Gamma(2H+3)}\,\frac{\left(A-A_{1}\right)^{2}}{t}\right]dA_{1}\,.
\ee
In the regime $A(t) \sim t^\beta$, the two terms in the exponent are of the same order of magnitude. 
Changing the integration variable $\sqrt{\frac{2}{c\,\Gamma(2H+3)}}\,r^{H+1-\beta}\,\frac{A_{1}}{t^{\beta}}\to z$ (and ignoring the Jacobian of this transformation as it is a subleading prefactor), we rewrite Eq.~\eqref{PrAtIntegralOverA1} as
\bea
\label{PrAtIntegralOverz}
P_{r}\left(A|t\right) &\sim&\int_{-\infty}^{\infty}\exp\left[-\left(rt\right)^{\alpha}\mathfrak{S}\left(y,z\right)\right]dz,\\
\label{yzdef}
y &=& \sqrt{\frac{2}{c\,\Gamma(2H+3)}}\,r^{H+1-\beta}\,\frac{A}{t^{\beta}},\quad
z = \sqrt{\frac{2}{c\,\Gamma(2H+3)}}\,r^{H+1-\beta}\,\frac{A_{1}}{t^{\beta}},\\
\label{mathfrakSdef}
\mathfrak{S}\left(y,z\right)&=&\nu\left(H\right)z^{2/\left(2H+3\right)}+\frac{1}{8}\left(y-z\right)^{2} \, ,
\eea
where
\be
\label{nudef}
\nu\left(H\right)=\frac{2^{-\frac{2}{2H+3}-1}(2H+3)((H+1)\Gamma(2H+3))^{\frac{1}{2H+3}}}{H+1} \, .
\ee
In the limit $t \to \infty$ with fixed $y$, we evaluate the integral \eqref{PrAtIntegralOverz} via the saddle-point approximation, the result recovering the anomalous scaling \eqref{larget.7}, but with a different representation of the rate function:
\be
\label{ratef.alternative}
\Phi\left(y\right)=\min_{z\in\left[0,y\right]}\left[\nu\left(H\right)z^{2/\left(2H+3\right)}+\frac{1}{8}\left(y-z\right)^{2}\right]\,.
\ee

We show the equivalence between the two representations for $\Phi(y)$ in Eqs.~\eqref{ratef.1} and \eqref{ratef.alternative} in Appendix \ref{app:PhiEquivalence}. Note that a very similar equivalence was shown in \cite{MGM21} for two representations of their rate function that coincide, up to scaling factors, with our $\Phi(y)$ with Hurst exponent $H=0$.
An advantage of the representation \eqref{ratef.alternative} is that it gives a clearer picture of the physical mechanism behind the condensation. The $z$ that is the minimizer in Eq.~\eqref{ratef.alternative} has the physical meaning: It gives the area under the condensate $A_{c}=zA/y$.
The duration of the run in which the condensate occurs is given by Eq.~\eqref{tau1star} with the replacement $A_1 \to A_c$. 
In the homogeneous phase, $y<y_c$, the minimizer in Eq.~\eqref{ratef.alternative} is $z=0$ and therefore $A_c$ vanishes. In contrast, deep into the condensed phase $A(t) \gg t^{\beta}$, we find $z \simeq y$, i.e., $A_{c} \simeq A$, so that in the leading order, $P_r(A|t)$ coincides with the distribution of the area under the first run, $P_r(A|t) \sim P_r(A_1)$ (this is somewhat similar to the ``big jump principle'' which occurs in large deviations of sums of i.i.d. random variables whose PDF decays slower than an exponential \cite{Chistyakov64,Foss13,Denisov08,Geluk09,Clusel06, BCV10,BUV14,VBB19, WVBB19,Gradenigo13,Barkai20, MKB98, BBBJ2000, EH05, MEZ05, EMZ06, Majumdar10, CC12, ZCG, Corberi15}).
The fact that the approximation $z \simeq y$ improves as $y$ is increased strongly suggests that the approximation $P_r(A|t) \sim P_r(A_1)$  holds even for very large $A$. In the next subsection, we use this argument in order to uncover a regime of very large fluctuations, $A(t) \sim t^{(2H+3)/2}$ in which a second-order dynamical phase transition occurs.
 Note that a similar analysis to that of the present subsection was also done recently for the area under a Ornstein-Uhlenbeck
process \cite{Smith22OU}.
Finally, in Appendix \ref{app:singleCondensate} we show that the assumption that we made shortly before Eq.~\eqref{PrAtIntegralOverA1} is consistent with the result \eqref{ratef.alternative}, which, in particular, means that in the supercritical regime $y>y_c$, a single condensate is optimal (i.e., far more probable than multiple condensates).

\subsection{Very large deviations}
\label{sec:veryLargeDeviations}

\begin{figure*}[ht]
\centering
\includegraphics[angle=0,width=0.46\textwidth]{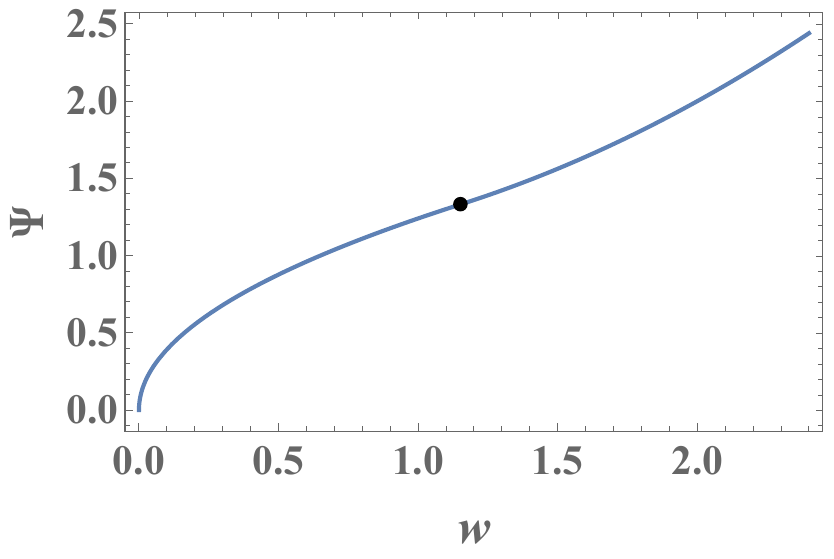}
\caption{The function $\Psi(w)$ for Brownian motion, $H=1/2$, see Eq.~\eqref{LDPPsi}. The critical point, corresponding to $w=w_c = 2/\sqrt{3}=1.1547\dots$, is marked by the black dot. At the critical point the second derivative of $\Psi(w)$ jumps, corresponding to a second-order dynamical phase transition.}
\label{FigPsi}
\end{figure*}

As described above, we argue that the coincidence $P_r(A|t) \sim P_r(A_1)$ persists even in the regime of very large fluctuations, $A(t) \sim t^{(2H+3)/2}$.
We therefore need to calculate $P_r(A_1)$ for $A_1 \sim t^{(2H+3)/2}$.
For $\frac{A_{1}}{\sqrt{cr}t^{\left(2H+3\right)/2}}<\sqrt{\frac{2}{H+1}}$, Eq.~\eqref{PrA1NearTail} holds but for $\frac{A_{1}}{\sqrt{cr}t^{\left(2H+3\right)/2}}>\sqrt{\frac{2}{H+1}}$ it does not,
because $\tau_1^*$ from Eq.~\eqref{tau1star} is larger than $t$. Thus, the minimizer of $\mathcal{S}\left(\tau_{1}\right)$ is $\tau_1=t$, and so the two terms in \eqref{PrA1Exact} give equal contributions (in the leading order of the saddle-point approximation that we use here), leading to a Gaussian decay of the $A_1\to\infty$ tail:
\be
\label{PrA1FarTail}
P_{r}\left(A_{1}\right)\sim e^{-\mathcal{S}\left(t\right)}\sim\exp\left[-\frac{A_{1}^{2}}{4ct^{2\left(H+1\right)}}-rt\right],\quad \frac{A_{1}}{\sqrt{cr}t^{\left(2H+3\right)/2}}>\sqrt{\frac{2}{H+1}}\,.
\ee
Now, using $P_r(A|t) \sim P_r(A_1)$, we find that $P_r(A|t)$ is also given by Eqs.~\eqref{PrA1NearTail} and \eqref{PrA1FarTail} (with the replacement $A_1 \to A$) which are conveniently written in the form of the anomalous LDP
\be
\label{LDPPsi}
P_{r}\left(A|t\right)\sim\exp\left[-rt\Psi\left(\frac{A}{\sqrt{cr}t^{\left(2H+3\right)/2}}\right)\right],\quad\Psi\left(w\right)=\begin{cases}
\frac{2H+3}{2H+2}\left(\frac{H+1}{2}\right)^{1/\left(2H+3\right)}w^{2/\left(2H+3\right)}\,, & w<\sqrt{\frac{2}{H+1}}\,,\\[3mm]
\frac{1}{4}w^{2}+1\,, & w>\sqrt{\frac{2}{H+1}}\,.
\end{cases}
\ee
The rate function $\Psi(w)$ is plotted in Fig.~\ref{FigPsi} for the case of Brownian motion, $H=1/2$.
Interestingly, it exhibits a second-order dynamical phase transition at the critical value $w_c = \sqrt{2/(H+1)}$, i.e., its second derivative jumps at $w=w_c$. The asymptotic behaviors near the critical point are
\be
\Psi(w)=\frac{2H+3}{2H+2}+\frac{w-\frac{\sqrt{2}}{\sqrt{H+1}}}{\sqrt{2}\sqrt{H+1}}+\begin{cases}
\frac{-(2H+1)\left(w-\frac{\sqrt{2}}{\sqrt{H+1}}\right)^{2}}{8H+12}+\dots, & w<\frac{\sqrt{2}}{\sqrt{H+1}}\,,\\[3mm]
\frac{1}{4}\left(w-\frac{\sqrt{2}}{\sqrt{H+1}}\right)^{2}+\dots, & w>\frac{\sqrt{2}}{\sqrt{H+1}}\,.
\end{cases}
\ee
In the subcritical regime $w < \sqrt{2/(H+1)}$, the prediction of \eqref{LDPPsi} is a stretched exponential, coinciding exactly with the tail \eqref{JointRegime} of the moderately-large-fluctuation regime. 
Therefore, the two large-deviation regimes have a joint regime of validity,
$t^{\beta}\ll A\left(t\right)<\sqrt{\frac{2cr}{H+1}}t^{\left(2H+3\right)/2}$,
 in which the distribution is given by Eq.~\eqref{JointRegime}.
The supercritical regime of $\Psi(w)$ describes a Gaussian decay of the distribution $P_r(A|t)$ as $A \to \infty$. However, note that this is completely different to the Gaussian distribution of typical fluctuations, Eq.~\eqref{PrAtGaussian}.

The phase transition of $\Psi(w)$ is qualitatively similar to the second-order transition found in \cite{MSS15a} when considering large deviations of the position of an RBM at finite time (see also \cite{SSIM21} and the very recent work \cite{SGS22} in which a similar phenomenon was found in subdiffusive resetting systems). One difference between the two cases is that in \cite{MSS15a} in the subcritical regime, the run that creates the fluctuation must be the last one ($i=n$) whereas in the present case, in the subcritical regime $w < w_c$, it can be any one of the runs $i\in\left\{ 1,\dots,n\right\}$.
The transition seperates between a phase $w>w_c$ in which it is created over the entire dynamics $\tau\in\left[0,t\right]$ and a phase
$w<w_c$ in which the fluctuation is created over a (strict) subinterval of the $\left[0,t\right]$.
In this respect, the transition is similar to those found in several systems without resetting \cite{SmithMeerson2019, MeersonSmith2019, AiryDistribution20, Meerson20}.

\section{Cumulants at late times}
\label{sec:cumulants}

We consider a Brownian motion with diffusion constant $D$, starting at
the origin and resetting to the origin at a constant rate $r$.
We are interested in extracting the late time behavior
of the cumulants of the area $A$ under the resetting Brownian motion (RBM)
of duration $t$, i.e., in this section we focus on the case $H=1/2$.
We start with the exact Fourier-Laplace transform
of the PDF $P_r(A|t)$ in Eq.~\eqref{main_res.1}
which we rewrite as
\begin{equation}
\int_{-\infty}^{\infty} dk\, e^{i\, k\, A}\, 
\int_0^{\infty} dt\, e^{-s\, t} \, P_r(A|t)=
\int_0^{\infty} dt\, e^{-st}\, \Big\langle e^{i\, k\, A}\Big\rangle
=\frac{{\tilde P}_0(k, s+r)}{1-r\, {\tilde P}_0(k, s+r)}\, ,
\label{exact.1}
\end{equation}
where for RBM (for which we recall $H=1/2$ and $c=D/3$), Eq.~\eqref{P0ks.1} becomes
\begin{equation}
{\tilde P}_0(k,s)= \int_0^{\infty} d\tau\, e^{-s\,\tau- \frac{D}{3}\, k^2\, 
\tau^3}\,.
\label{P0ks.2}
\end{equation}

We recall the definition of the cumulants of a random variable 
\begin{equation}
\Big\langle e^{i\, k\, A}\Big\rangle= \exp\left[ \sum_{n=1}^{\infty} \frac{(ik)^n}{n!}\, \langle A^n\rangle_c\right]\, ,
\label{cum_def.1}
\end{equation}
where $\langle A^n\rangle_c$ is the $n$-th cumulant. 
The cumulants of a sum of i.i.d. random variables are exactly proportional to the number of terms in the sum, and it is natural to expect this behavior to extend to continuous-time systems for dynamical observables \eqref{dynamicalObservable} in the long-time limit, when $t$ is much larger than the typical correlation time of the system. Therefore, we anticipate that
at late times $t$, the cumulants will all scale linearly with $t$, i.e.,
\begin{equation}
\langle A^n\rangle_c\approx c_n \, t\, .
\label{cum_t.1}
\end{equation}
Our goal is to extract the coefficients $c_n$'s. Using the anticipated scaling
in Eq. (\ref{cum_t.1}) in Eq. (\ref{cum_def.1}), we expect that at late times
\begin{equation}
\Big\langle e^{i\, k\, A}\Big\rangle \approx e^{b(k)\, t}\, ; \quad {\rm where}\quad 
b(k)= \sum_{n=1}^{\infty} \frac{(ik)^n}{n!}\, c_n\, .
\label{char.1}
\end{equation}
Substituting this anticipated late time behavior on the left hand side (lhs)
of Eq. (\ref{exact.1}), we get
$\int_0^{\infty} dt\, e^{-(s- b(k))\, t}$ which clearly diverges
when $s= b(k)$. This indicates that the right hand side (rhs)
of Eq. (\ref{exact.1}) must have a pole at $s=b(k)$. In other words,
\begin{equation} 
1-r\, {\tilde P}_0(k, b(k)+r)=0\, .
\label{pole.1}
\end{equation}

To simplify a bit, let us further define
\begin{equation}
a(k)= 1+\frac{1}{r}\, b(k)= 1+\frac{1}{r}\sum_{n=1}^{\infty} \frac{(ik)^n}{n!}\, c_n\, .
\label{def_ak}
\end{equation}
Then Eq. (\ref{pole.1}) can be rewritten as (upon using Eq. (\ref{P0ks.2}) and
rescaling $r\tau\to \tau$)
\begin{equation}
\int_0^{\infty} d\tau\, e^{-a(k)\, \tau - \frac{Dk^2}{3 r^3}\, \tau^3}=1\, .
\label{pole.2}
\end{equation} 
For each $k$, one needs to find the positive root of this transcendental
equation to obtain $a(k)$ and once we know the power series expansion
of $a(k)$, we can read off $c_n$ from it using (\ref{def_ak}).
Let us first remark that from Eq. (\ref{pole.2}) it is clear
that $a(k)$ is only a function of $k^2$, indicating that
all odd cumulants vanish as expected, i.e.,
\begin{equation}
a(k)= 1+ \frac{1}{r} \sum_{m=1}^{\infty} \frac{(-1)^m k^{2m}}{(2m)!}\, c_{2m}\, .
\label{ak.2}
\end{equation}
To make further progress, let us expand $e^{-D\,k^2\, \tau^3/(3r^3)}$ in powers of $k^2$
and perform the resulting integration over $\tau$ in Eq. (\ref{pole.2}) term by term.
This gives
\begin{equation}
\sum_{n=0}^{\infty} \frac{(-1)^n (3n)!}{n!}\, \left(\frac{Dk^2}{3 r^3}\right)^n 
\frac{1}{\left[a(k)\right]^{3n+1}}= 1\, .
\label{pole.3}
\end{equation} 
We can now substitute the power series expansion of $a(k)$ in (\ref{ak.2})
on the lhs of (\ref{pole.3}) and finally expand the lhs as a power series in $k^2$.

This gives, for instance, up to order $k^4$
\begin{equation}
\left(\frac{c_2}{2r}-\frac{2D}{r^3}\right)\, k^2 
+ \left(\frac{c_2^2}{4r^2}-\frac{c_4}{24r}- \frac{4Dc_2}{r^4}+\frac{40D^2}{r^6}\right)\, k^4 +O(k^6)=0\, .
\label{pole.4}
\end{equation}
Since this holds for all $k$, the coefficients must all vanish which then allows
to determine $c_n$'s recursively. 
From Eq.~\eqref{pole.4}, we get the first two nonvanishing coefficients
\begin{eqnarray}
c_2 & = & \frac{4D}{r^2}\,, \label{c2.1} \\
c_4 &=  &\frac{6}{r}c_2^2 - \frac{96 D}{r^3} c_2 + \frac{960 D^2}{r^5}= \frac{672\, D^2}{r^5}\, .
\label{c4.1}
\end{eqnarray}
This result can be compared with that of Ref.~\cite{DH2019}, where the second and fourth moments were calculated exactly. For $D=1/2$, they found
\bea
\label{moment2}
\left\langle \left[A\left(t\right)\right]^{2}\right\rangle &=&\frac{2}{r^{3}}\left[rt-2+e^{-rt}\left(2+rt\right)\right] \,,\\
\left\langle \left[A\left(t\right)\right]^{4}\right\rangle &=&\frac{1}{r^{6}}\left\{ 12\left(rt\right)^{2}+120rt-840+e^{-rt}\left[9\left(rt\right)^{4}+68\left(rt\right)^{3}+288\left(rt\right)^{2}+720rt+840\right]\right\}\,. 
\eea
The second cumulant equals the second moment. The long-time limit $r t \gg 1$ in \eqref{moment2} then yields $\left\langle A^{2}\right\rangle _{c}\simeq2t/r^{2}$ in agreement with our \eqref{c2.1} with $D=1/2$.
The fourth cumulant is exactly given, in terms of the second and fourth moments, by
\bea
\left\langle A^{4}\right\rangle _{c}&=&\left\langle \left[A\left(t\right)\right]^{4}\right\rangle -3\left(\left\langle \left[A\left(t\right)\right]^{2}\right\rangle \right)^{2} \nn\\
&=&\frac{168t}{r^{5}}-\frac{888}{r^{6}}+e^{-rt}\left(\frac{936}{r^{6}}+\frac{720t}{r^{5}}+\frac{264t^{2}}{r^{4}}+\frac{68t^{3}}{r^{3}}+\frac{9t^{4}}{r^{2}}\right)+e^{-2rt}\left(-\frac{48}{r^{6}}-\frac{48t}{r^{5}}-\frac{12t^{2}}{r^{4}}\right) \, .
\eea
In the long-time limit, the leading order term $\left\langle A^{4}\right\rangle _{c}\simeq168t/r^{5}$ indeed agrees with our \eqref{c4.1} with $D=1/2$.

Using Mathematica, we extended the procedure described above in order to calculate the lowest nonvanishing coefficients up to $c_{12}$.
Expanding Eq.~(\ref{pole.3}) in powers of $k^2$ using the $a(k)$'s from (\ref{ak.2}), 
we found, for $D=r=1$,
\begin{equation}
a(k) = 1-2 k^2+28 k^4-1640 k^6+194064 k^8-37369984 k^{10}+10566455104 k^{12} + \dots
\end{equation}
which gives us, in addition to $c_2$ and $c_4$ that are given above,
\begin{equation}
\label{cumulantsUpTo12}
c_{6}=\frac{1180800D^{3}}{r^{8}},\quad c_{8}=\frac{7824660480D^{4}}{r^{11}},\quad c_{10}=\frac{135608197939200D^{5}}{r^{14}},\quad c_{12}=\frac{5061348901144166400D^{6}}{r^{17}}\,,
\end{equation}
where the $D$ and $r$ dependence was easily restored because $c_{n}\propto D^{n/2}r^{-3n/2+1}$ from dimensional analysis.
The coefficients $c_n$ can clearly be seen to grow very rapidly with $n$. Accordingly,
$a(k)$ should be interpreted as a formal power series: It does not define a function of $k$ since the sum \eqref{ak.2} diverges for any nonzero $k$.

 The same method that we used in this section can be 
 extended to general $H$, but we will not pursue this here.

\section{Summary and discussion}  
\label{sec:conclusion}

To summarize, we calculated the full distribution of the area under an SGP with stochastic resetting at long times $t\to\infty$. The usual large-deviation scaling \eqref{eq:DVscaling} does not hold.
Instead, we uncovered two anomalous LDPs for two different large-deviation regimes, and calculated the exact rate functions for each regime. Moreover, we found that each of the two rate functions $\Phi(y)$ and $\Psi(w)$ has a singularity, corresponding to dynamical phase transitions of the first and second order, respectively.
The transition in $\Phi(y)$ is of a condensation type, and remarkably, $\Phi(y)$ coincides, up to scaling factors, with rate functions found in other systems in which condensation transitions occur \cite{GM19, BKLP20, GIL21, GIL21b, MGM21, MLMS21, Smith22OU},
such as the run-and-tumble particle, nonlinear breathers,
Ornstein-Uhlenbeck process etc. All these problems share
a common feature that one is interested effectively in the sum of
a number of IID random variables, $A=\sum_{i=1}^n {A_i}$.
The condensation transition occurs when the sum includes a single term $A_i$ that is macrosopic, i.e., of the same order as the entire sum $A$.
It turns out that the criterion for this transition
is the following \cite{MLMS21}: when the distribution of the
underlying
random variables has a stretched exponential tail, $p(A)\sim\exp\left(-|A|^{a}\right)$
with the stretching
exponent $ 0<a<1$, then condensation occurs, accompanied
by an anomalous large deviation form.
In our problem, there are two aspects (i) the number $n$ of
random variables involved in the sum $A=\sum_{i=1}^n A_i$
is random  and (ii) Each of them has a stretched exponential
tail with $0<a= 2/(2H+3)<1$ (Eq.~\eqref{PrA1NearTail}), that also leads to
a condensation transition with an anomlaous large deviation form.
In our problem, there is the additional feature that at very large areas, Eq.~\eqref{PrA1NearTail} breaks down and gives way to Eq.~\eqref{PrA1FarTail}. This is what leads to the existence of the very-large-fluctuations regime and to the dynamical phase transition in $\Psi(w)$, which is qualitatively similar to transitions which have been observed in other contexts \cite{MSS15a}. 

We found that, despite the anomalous scaling of the full distribution, its cumulants grow (asymptotically) linearly
in time  and developed a method for calculating the coefficients.
Remarkably, and in contrast to the case in which the full distribution follows the ``usual'' LDP \eqref{eq:DVscaling}, there is no obvious connection between cumulants and large deviations in our system.
This is because the anomalous LDP \eqref{larget.7} doesn't give any corrections to the Gaussian distribution in the typical fluctuations regime $A(t)\sim\sqrt{t}$ because the rate function $\Phi(y)$ is exactly parabolic around its minimum at $y=0$.
We expect these features to be universal for a broader class of systems in which condensation transitions occur \cite{NT18, GM19, BKLP20, GIL21, GIL21b, MGM21, MLMS21, Smith22OU}. Indeed, the cumulants of a sum $\sum_{i=1}^{n}x_{i}$ of i.i.d. random variables $x_1, \dots, x_n$, are exactly proportional to $n$, and it is  natural to expect this behavior to extend to continuous-time systems for dynamical observables \eqref{dynamicalObservable} in the long-time limit, when $t$ is much larger than the typical correlation time of the system.
Interestingly, there are systems in which the behavior is exactly opposite to that observed here in the sense that the cumulants grow anomalously in time, while the scaling \eqref{eq:DVscaling} is not found to break down. Such behaviors were found in the two recent works \cite{Krajnik21,Krajnik22}, and it would be interesting to investigate whether these different anomalous behaviors are related.

The two key ingredients in the analysis performed in subsection \ref{sec:condensedPhase} in which, in particular, the moderately-large-fluctuations result was reproduced, were the Gaussian behavior of typical fluctuations \eqref{PrAtGaussian} and the near tail of the area under a single run \eqref{PrA1NearTail}. 
Similarly, in subsection \ref{sec:veryLargeDeviations} only the near \eqref{PrA1NearTail} and far \eqref{PrA1FarTail} tails of the area under a single run were needed.
Note that these analyses did not directly use the exact result \eqref{main_res.2} and can therefore be applied in a much broader range of scenarios, even when exact results are unavailable.
For example, the absolute area
$B(t) = \int_{0}^{t}|x\left(\tau\right)|d\tau$
was studied in \cite{DH2019} for the RBM. It was found that the usual scaling \eqref{eq:DVscaling} holds for $B(t)$ smaller than its mean, $B\left(t\right)<\left\langle B\left(t\right)\right\rangle$,   and the corresponding rate function $I(a)$ was calculated, but the full distribution for $B(t)$ has remained unknown for $B\left(t\right)>\left\langle B\left(t\right)\right\rangle $, and we now outline its calculation.
The dominant contribution to the $A_1 \to +\infty$ tail of the area under a single run comes from realizations in which $x(\tau)$ is positive \cite{MO22}, so Eqs.~\eqref{PrA1NearTail} and \eqref{PrA1FarTail} should extend to the absolute area under a single run as well, while typical fluctuations will still follow a Gaussian distribution with variance $\sim t$, but with a different coefficient to the one in Eq.~\eqref{PrAtGaussian}.
As a result,
an analysis analogous to that of subsections \ref{sec:condensedPhase} and \ref{sec:veryLargeDeviations} for the absolute area would show that the distribution in the regime $B\left(t\right)>\left\langle B\left(t\right)\right\rangle$ follows the same anomalous LDPs as the area $A(t)$, i.e., it is given by Eqs.~\eqref{larget.7} and \eqref{LDPPsi} (replacing $A \to B$). The rate function $\Phi(y)$ would still be given by Eq.~\eqref{ratef.alternative} only with a different numerical coefficient instead of the coefficient $1/8$ (due to the different variance), while $\Psi(w)$ would remain unchanged.

 Another extension would be to the fixed-$n$ ensemble, where $n-1$ is the number of resetting events. In the fixed-$n$ ensemble, the areas $A_1, \dots, A_n$ under the runs become i.i.d. random variables whose distribution tails decay slower then exponentially, and are given by Eq.~\eqref{PrA1NearTail}. Thus, based on the general discussions in Refs. \cite{BKLP20, MLMS21}, we expect the moderately-large-fluctuation regime to exhibit very similar behavior to that of the fixed-$t$ ensemble studied here, including the condensation transition.
In contrast, the very-large-fluctuation regime should be absent in the fixed-$n$ ensemble, because its very existence is the result of a finite-$t$ effect.
From a more general point of view, we expect a similar condensation phenomenon to occur in the large deviations of dynamical observables in other stochastically-resetting systems as long the tail of the distribution of the observable under a single run decays slower then exponentially. This could occur with other resetting protocols too.

\acknowledgements

SNM acknowledges support from the ANR grant ANR-17-CE30- 0027-01 RaMaTraF.

\begin{appendices}

\section{Equivalence of the two representations for the rate function $\Phi(y)$}
\label{app:PhiEquivalence}

\renewcommand{\theequation}{A\arabic{equation}}
\setcounter{equation}{0}

In the main text, we found two representations for the rate function $\Phi(y)$ that describes moderately large deviations: Eqs.~\eqref{ratef.1} and \eqref{ratef.alternative}. The goal of this appendix is to show that the two representations are equivalent.
The representation \eqref{ratef.1} can be rewritten as 
\be
\label{Phichi}
\Phi\left(y\right)=\min\left\{ \frac{y^{2}}{8},\chi\left(y\right)\right\}
\ee
where $\chi(y)$ is given parametrically by Eqs.~\eqref{chi.1} and \eqref{chi.2}.
Similarly, the representation \eqref{ratef.alternative} can also be written in the form \eqref{Phichi} 
with the function $\chi(y)$ that is given parametrically by
\bea
\label{chiAlternative1}
\chi&=&\nu\left(H\right)z^{2/\left(2H+3\right)}+\frac{1}{8}\left[\frac{8\nu\left(H\right)}{2H+3}z^{-\left(2H+1\right)/\left(2H+3\right)}\right]^{2} \, ,\\
\label{chiAlternative2}
y&=&\frac{8\nu\left(H\right)}{2H+3}z^{-\left(2H+1\right)/\left(2H+3\right)}+z \, .
\eea
To reach Eq.~\eqref{chiAlternative2}, one solves the equation $\partial\mathfrak{S}/\partial z=0$ for $y$, where $\mathfrak{S}(y,z)$ is defined in \eqref{mathfrakSdef}. Eq.~\eqref{chiAlternative1} is then reached by plugging Eq.~\eqref{chiAlternative2} into \eqref{mathfrakSdef}.
Thus, in order to show the equivalence between the two representations for $\Phi(y)$, it is sufficient to show 
that the two representations for $\chi(y)$ are equivalent.
Indeed, one finds that by plugging 
\be
u=\left(\frac{2H+3}{8\nu\left(H\right)}\right)^{1/\left(2H+2\right)}z^{2/\left(2H+3\right)}
\ee
into the representation \eqref{chi.1} and \eqref{chi.2}, one obtains the representation \eqref{chiAlternative1} and \eqref{chiAlternative2}.

\section{Optimality of a single condensate}
\label{app:singleCondensate}

\renewcommand{\theequation}{B\arabic{equation}}
\setcounter{equation}{0}

In this appendix, we show that in the supercritical regime $y>y_c$, a single condensate is optimal (i.e., far more probable than multiple condensates).
We do this by checking that the assumption that we made shortly before Eq.~\eqref{PrAtIntegralOverA1} is consistent with the result \eqref{ratef.alternative}, and therefore the area $A-A_1$ is sufficiently small so that it is not worthwhile for the system to create a second condensate.
So, we must show that $A-A_1$ is in the subcritical regime, i.e., that
\be
\sqrt{\frac{2}{c\,\Gamma(2H+3)}}\,r^{H+1-\beta}\,\frac{A-A_{1}}{t^{\beta}}<y_{c},
\ee
or equivalently, using Eq.~\eqref{yzdef}, that $y-z_1<y_{c}$ where $z_1$ is the minimizer in Eq.~\eqref{ratef.alternative}, i.e., $\Phi(y) = \mathfrak{S}\left(y,z_{1}\right)$.
Let us assume that this is not the case, i.e., that $y-z_1>y_{c}$. This would imply that when calculating $\Phi(y-z_1)$ via Eq.~\eqref{ratef.alternative}, the minimum is obtained at some $z = z_2 \ne 0$, i.e., that 
\be
\mathfrak{S}\left(y-z_{1},z_{2}\right)=\nu\left(H\right)z_{2}^{2/\left(2H+3\right)}+\frac{1}{8}\left(y-z_{1}-z_{2}\right)^{2}<\frac{1}{8}\left(y-z_{1}\right)^{2}=\mathfrak{S}\left(y-z_{1},0\right) \, .
\ee
However, this would in turn lead to.
\bea
\label{yz1z2}
\mathfrak{S}\left(y,z_{1}\right)&=&\nu\left(H\right)z_{1}^{2/\left(2H+3\right)}+\frac{1}{8}\left(y-z_{1}\right)^{2}>\nu\left(H\right)z_{1}^{2/\left(2H+3\right)}+\nu\left(H\right)z_{2}^{2/\left(2H+3\right)}+\frac{1}{8}\left(y-z_{1}-z_{2}\right)^{2} \nn\\
&>&\nu\left(H\right)\left(z_{1}+z_{2}\right)^{2/\left(2H+3\right)}+\frac{1}{8}\left(y-z_{1}-z_{2}\right)^{2}=\mathfrak{S}\left(y,z_{1}+z_{2}\right) \, ,
\eea
in contradiction with the minimality $\forall z,\;\Phi\left(y\right)=\mathfrak{S}\left(y,z_{1}\right)\le\mathfrak{S}\left(y,z\right)$ of $z_1$.
Note that in Eq.~\eqref{yz1z2}, when moving from the first to the second, we used the concavity of the function $z \to z^{2/\left(2H+3\right)}$ (which holds since we assume $H>0$ throughout the paper).

\end{appendices}

 \end{document}